\documentclass[%
 reprint,
 amsmath,amssymb,
 aps,
]{revtex4-2}

\usepackage{graphicx}% Include figure files
\usepackage{dcolumn}% Align table columns on decimal point
\usepackage{bm}% bold math
\usepackage{subcaption}
\usepackage{tabularx,booktabs} %Packages for the tables
\newcolumntype{Y}{>{\centering\arraybackslash}X} %Definisco il tipo di colonna Y all'interno dell'ambiente tabular, è uguale al tipo X ma è centrato
\usepackage{array}
\usepackage{caption}
\usepackage{hyperref}% add hypertext capabilities
\begin{document}

\preprint{APS/123-QED}

\title{A new class of axion haloscope resonators:\\the polygonal coaxial cavity}% Force line breaks with \\

\author{R.~Di Vora}
\email[The authors to whom correspondence may be addressed: \\]{divora@lnl.infn.it}
\affiliation{INFN, Laboratori Nazionali di Legnaro, Legnaro, Padova, Italy}
\author{C.~Braggio}
\affiliation{Dipartimento di Fisica e Astronomia, Padova, Italy}\affiliation{INFN, Sezione di Padova, Padova, Italy} 
\author{G.~Carugno} \affiliation{INFN, Sezione di Padova, Padova, Italy}
\author{A.~Gardikiotis} \affiliation{INFN, Sezione di Padova, Padova, Italy}
\author{A.~Lombardi} \affiliation{INFN, Laboratori Nazionali di Legnaro, Legnaro, Padova, Italy}
\author{A.~Ortolan} \affiliation{INFN, Laboratori Nazionali di Legnaro, Legnaro, Padova, Italy}
\author{G.~Ruoso} \affiliation{INFN, Laboratori Nazionali di Legnaro, Legnaro, Padova, Italy}

%\affiliation{%

\date{\today}% It is always \today, today,

\begin{abstract}
%Abstract: novel high effective volume and high tunability cavity for axion searches, based on ... design; main advantages/characteristics of the design, summary of simulation results and experimental tests (bead pulling [?], tunability)

In the search for axionic Dark Matter, the high frequency part of the QCD axion parameter space is favored, as indicated by both cosmological and astrophysical arguments and recent indications from lattice QCD calculations.
To extend the probing range of cavity haloscopes, solutions addressing the unfavorable scaling of cavity volume with frequency must be developed. %Therefore there is great interest in developing solutions
Here, we present a novel type of high-volume thin shell resonator for high frequency haloscope dark matter searches. The cavity is formed by two nested and coaxial right angle polygonal prisms enclosed within two flat endcaps. For the axion-sensitive (pseudo-)TM010 mode, finite element simulations yield form factor of the order of 0.8 and Q factor of the order of 60000 for a copper cavity at 4\,K. High tunability of up to $\sim 5\%$ is achieved by reciprocal rotation of the two prisms, without significant changes in haloscope sensitivity. A prototype aluminium hexagonal cavity was built and tested, confirming the main characteristics of the design.

\end{abstract}

\maketitle

%\tableofcontents

\section{\label{sec:intro}Introduction}%\protect\\ The line break was forced \lowercase{via} \textbackslash\textbackslash}

%Brief resume of the state of the field: high-Tc superconductive cavities (CAPP), multiple cavities (CAST, CAST-CAPP), dielectrical cavities (QUAX), coaxial conical cavity (Taiwan), [tunable cavity at QUAX?]. plasma haloscopes

The axion is a pseudo-Goldstone boson introduced by Weinberg\,\cite{PhysRevLett.40.223} and Wilczek\,\cite{PhysRevLett.40.279} as part of the Peccei-Quinn mechanism to solve the Standard Model puzzle known as the strong-CP problem of Quantum Chromodynamics (QCD)\,\cite{PhysRevLett.38.1440,PhysRevD.16.1791}. In current axion models, the QCD axion would have low mass and very weak interaction with ordinary matter. Additionally, cosmological axions created during the Peccei-Quinn phase transition would be non-relativistic and have decay times much longer than the age of the universe, making such 'invisible' axions ideal candidates for the composition of Dark Matter. 
Astrophysical bounds and cosmological considerations select a range of $1\,\mu$eV\,$ < m_a < 10$\,meV\,\cite{IRASTORZA201889,Workman:2022ynf}, with recent lattice QCD calculations favoring the 40-180\,$\mu$eV region\,\cite{Buschmann:2021sdq}.

In the last few years, increasing efforts have been invested into the detection of the axion as the possible majority constituent of Cold Dark Matter, with a large portion of the experiments employing the axion haloscope originally proposed by P.\,Sikivie\,\cite{PhysRevLett.51.1415}. 
In its simplest form, this experimental scheme exploits the large temporal (and spatial) coherence of the axion field by collecting the power released in a microwave resonator immersed in a strong magnetic field.
The converted axion power is (in the limit $Q_0\ll Q_a$):
\begin{equation}\label{eq:P_axion}
P_a = g_{a \gamma \gamma}^2 \frac{\rho_a}{m_a^2} \omega_c B_0^2 C V Q_0 \frac{\beta}{(1+\beta)^2} \; ,
\end{equation}
where $g_{a \gamma \gamma}$ is the axion to photon effective coupling, $\rho_a\,\sim\,0.45$\,GeV/cm$^3$ is the axion density\,\cite{Workman:2022ynf}, $m_a$ is the axion mass, $\omega_c$ is the cavity frequency, $B_0$ the employed static magnetic field%(the value of which depends on the chosen convention for the $C$ factor formula, see next paragraph)
, $Q_0$ is the unloaded quality factor of the cavity mode, and $\beta$ is the coupling coefficient of the antenna to the cavity mode. 

The form factor $C$ expresses the overlap between the electric field $\mathbf{E_{mnl}}(\vec{x})$ of the employed cavity mode and the external magnetic field $\mathbf{B}(\vec{x})$ enabling the axion-to-photon conversion. It is given by
\begin{equation}\label{eq:C_factor}
C=\frac{{\lvert} \int_V \, \mathbf{E_{mnl}}\cdot \mathbf{B} \; \mathrm{d^3}x \,{\rvert}^2}{\int_V \, {\lvert} \mathbf{B} {\rvert}^2 \, \mathrm{d^3}x \, \int_V \, \epsilon {\lvert} \mathbf{E_{mnl}} {\rvert}^2 \, \mathrm{d^3}x } \; ,
\end{equation}
where $\epsilon(\vec{x})$ is the dielectric constant at every point $\vec{x}$ in the cavity volume V.
%Note that due to the magnetic integral at the denominator serving as normalization, if we use this formula $B_0$ would be the quadratic mean of the magnetic field in the cavity. However, due to the custom of quoting the peak magnetic field provided by the solenoid as $B_0$, the $C$ factor is usually renormalized accordingly.
In the following we will assume $\mathbf{B}(\vec{x}) = B_0 \mathbf{\hat{z}}$.

Since the axion frequency is to date unknown, an axion haloscope must be able to tune its resonator to cover a range of frequencies. The speed at which this range can be scanned at a certain level of sensitivity in $g_{a\gamma\gamma}$ is called the haloscope scan rate\,\cite{Kim:2020kfo}. It is directly proportional to the so-called 'figure of merit' $F$, which collects all the cavity-related parameters
\begin{equation}\label{eq:fig_of_merit}
F=(CV)^2Q_0 \; ,
\end{equation}
the most useful quantity in the design of a new cavity.

\begin{figure*}[ht!]
    \centering
       \subcaptionbox{\small Electric field distribution as seen from eigenmode simulation.}{\includegraphics[width=.49\linewidth]{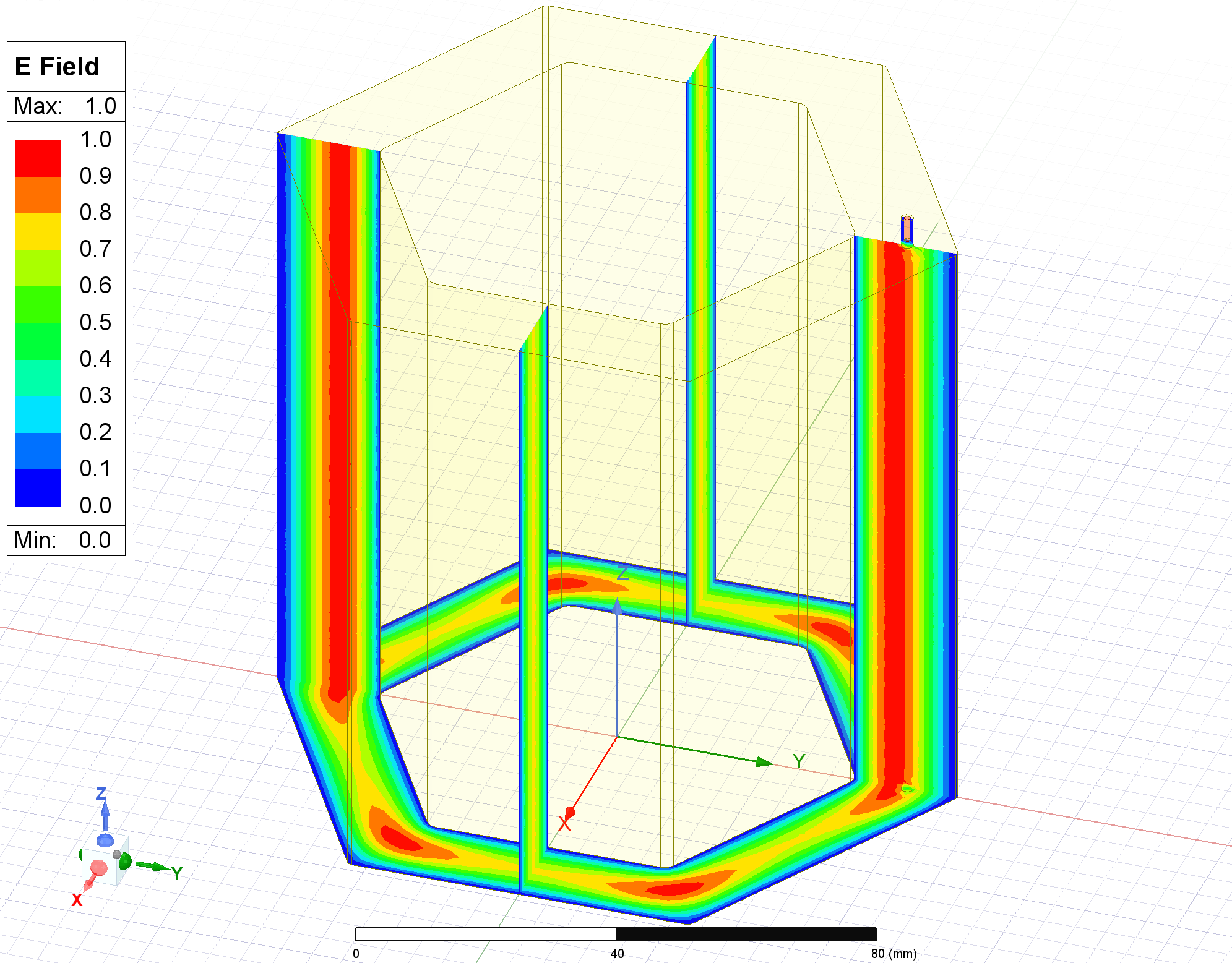}}\quad
       \subcaptionbox{\small Magnetic field distribution as seen from eigenmode simulation.}{\includegraphics[width=.49\linewidth]{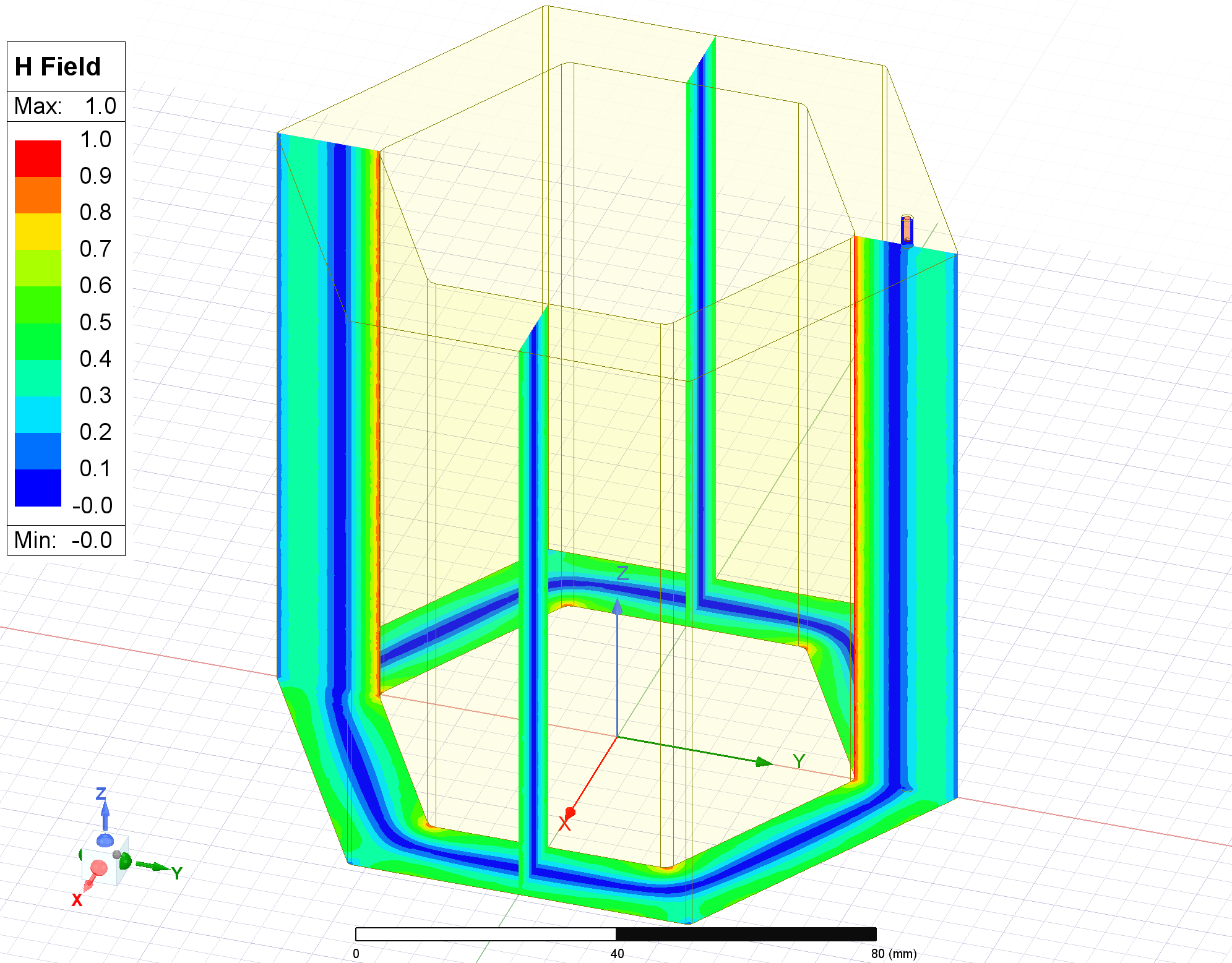}}\quad
\caption{\small \label{fig:Cav_model_E_B_distr} Basic cavity model used for simulation studies. The cavity is an empty region delimited by two solid copper coaxial polygonal prisms and two endcaps. Each endcap hosts a coax port used for modal simulations. The auxiliary copper structure enveloping the cavity is not shown in the figure.}
\end{figure*}

Solenoids are the common source of strong intensity magnetic fields. For a given coil length, the available volume scales with the square radius of the bore.
Assuming the use of a cylindrical empty cavity as the resonator, the figure of merit is maximized by its fundamental mode TM$_{010}$, whose resonant frequency is set by the inverse of the diameter of the cavity. An increase in the targeted mass then automatically results in a quadratic reduction of the volume of the cavity. 
The haloscope approach is then extremely effective towards the lower end of the QCD axion mass search range, with the most sensitive experiments having already been able to probe the theoretically expected couplings\cite{PhysRevD.103.032002,PhysRevLett.130.071002}. %Conversely, this characteristic inverse quadratic scaling becomes its main limitation at higher frequencies, limiting the cavity volume not only below the axion coherence length, but also well below the available magnet bore (which is the usual limiting factor at lower frequencies). 

%A moltitude of different haloscope designs are being explored with the aim of increasing the signal in such conditions, typically by attempting to increase the effective volume by various means. Among the most notable solutions, some use an array of identical cavities [NB! \,\cite{REF}], some involve the use of differently shaped cavities [NB! \,\cite{REF}] or of dielectrics to exploit higher order modes [NB! \,\cite{REF}], and finally some do away with the concept of a cavity altogether (i.e. the so-called 'plasma haloscope' [NB! \,\cite{REF}]). Some of these solutions are theoretically more effective than others, but all present different challenges to be overcome, mostly from a practical standpoint and in particular in the construction and operation of increasingly complicated and precise tuning mechanisms.

%It is in an attempt to mitigate this latter issue that we present in the following the design for a widely- and easily-tunable resonator for DM haloscopes. The employed design, while not able to fully exploit the full bore of the available magnet as in some of the previously introduced approaches\footnote{We note however, that given the findings reported in this work it should be possible to increase the bore occupancy by fitting multiple concentric cavities, possibly with independent tuning and/or different number of sides.}, still represents a decisive improvement on the standard TM$_{010}$ empty cylindrical cavity paradigm and exploits what we think is the most simple and robust tuning system possible, the rigid rotation around a single, fixed axis.

To address the unfavorable scaling of the signal towards high frequencies, several solutions were put forth to increase the occupation of the magnet bore. Among them, tunable multirod cavities\,\cite{Stern:2015kzo,10.1063/5.0016125}, dielectric cavities\,\cite{PhysRevApplied.9.014028,PhysRevApplied.14.044051,Kim:2019asb,QUAX:2020uxy,PhysRevApplied.17.054013,QUAX:2020wfd,PhysRevD.107.015012}, photonic band gap cavities (the so-called "plasma haloscope")\,\cite{Kishimoto:2021ral,PhysRevLett.123.141802,ALPHA:2022rxj}, the so-called "pizza cavity" or multi-cell concepts\,\cite{Jeong:2017hqs,Jeong:2022akg,AlvarezMelcon:2020vee}, and the use of an array of $K$ cavities\,\cite{10.1063/1.1141427,Kinion:2001fp,Yang:2020xsc,Adair:2022rtw,JEONG201833}. %Provided the challenge to match cavity frequencies and phase is overcome, 
The latter solution is particularly advantageous when the cavities are coherently readout, as the sensitivity gain is proportional to $K$, instead of just $\sqrt{K}$ as is the case of incoherent arrays\,\cite{kinion2001first,Adair:2022rtw,JEONG201833}. %Coherent readout of four cavities in a single magnet have been employed to search for DM axions\,\cite{kinion2001first,Adair:2022rtw}. %A coherent readout of four rectangular cavities inserted in a dipole magnet has also been employed to probe Axion-Like Particles (ALPs) in the $19.74\,\mu$eV to $22.47\,\mu$eV mass range\,\cite{Adair:2022rtw}. 
However, exploiting coherence becomes increasingly difficult with the number of cavities due to the high accuracy requirements of the mechanical tuning system. 
The pizza cavity concept instead relies on a single cylindrical cavity filling the available bore, where identical azimuthal slices are separated through conductive diaphragms. This approach, while interesting to maximize the effective volume, leads to a rapid increase of the number of partitions for larger magnet bore radius, further complicating the tuning system. 
Recently, a new approach has emerged in the form of folded thin shell cavities\,\cite{Kuo:2019cps,Kuo:2020llc,Dyson:2024elo}, where a large volume is obtained by folding a sheet-like cavity along one of the dimensions. This solution also allows for very wide tuning by a very precise variation of the shells separation.

In this work we introduce a new, widely- and easily- tunable high-volume resonator for high frequency DM haloscopes. It is based on two nested, polygonal coaxial prisms enclosed within two flat endcaps. %This design, while not able to exploit the bore of the available magnet fully as in some of the previously introduced approaches%\footnote{We note however, that given the findings reported in this work it should be possible to increase the bore occupancy by fitting multiple concentric cavities, possibly with independent tuning and/or different number of sides.}
This design represents a decisive improvement on the standard TM$_{010}$ empty cylindrical cavity paradigm and exploits what we think is the most simple and robust tuning system possible, i.e. the rigid rotation around a single, fixed axis.

In Sec.\,\ref{sec:design}, we describe the cavity design and its fundamental characteristics. Sec.\,\ref{sec:tuning} details the results of the simulation evaluating the cavity tuning range and the stability of the cavity parameters over this range. We show that as a consequence of the tradeoff between tuning range and effective volume, once operational frequency and external diameter are chosen, there exists an optimal choice of cavity design for each desired tuning range. 
In Sec.\,\ref{sec:tolerances} we present the results of indicative studies on the influence on the cavity parameters of nonidealities in the cavity geometry, leading to the determination of the maximum allowable tolerances.
Experimental results obtained for a full-scale prototype of the cavity are detailed in Sec.\,\ref{sec:measurements}. In Sec.\,\ref{sec:comparisons} the projected sensitivity of the studied approach is then compared to the benchmarks provided by pizza cavities and cavity arrays.
%The methods and techniques employed in simulation studies and esperimental measurements are reported in Sec.\,\ref{sec:methods}.
Finally, in Sec.\,\ref{sec:conclusions} we briefly recap the obtained results and draw our conclusions.

%NB! What about plasma haloscope? Just found this article with experimental results from japanese authors in 2021\,\cite{Kishimoto:2021ral}, and Millar\,\cite{PhysRevLett.123.141802},\,\cite{ALPHA:2022rxj}
%And the DBAS sapphire wedge cavity from Tobar\,\cite{PhysRevApplied.14.044051}?
%NB! Lì delle multicell, RADES fa divisione longitudinale quindi non va bene per cavità in solenoide largo (per uno lungo OK!), ma in magnete dipolare!

%NB! Thin shell cavities:\,\cite{Kuo:2019cps},\,\cite{Kuo:2020llc}
%Articolo di Tai Dyson ancora in arxiv su wedge cavity\,\cite{Dyson:2024elo}

\section{\label{sec:design}The tunable polygonal resonator}
In designing the cavity subject of this work, we first considered a thin shell cavity delimited by two nested coaxial and concentric cylinders of equal height and flat endcaps. The radii of the two cylinders are $r_{\rm in}$ and $r_{\rm ext}$ respectively (with $r_{\rm in}<r_{\rm ext}$). Given the cavity length $L$, we choose a gap $d=r_{\rm ext}-r_{\rm in}\ll \{L,\,r_{\rm in}\}$. The properties of such a cavity are easily derived in the limit of large inner radius and infinite sides, where it simplifies to an infinite rectangular box. The gap $d$ sets the frequency of the TM$_{010}$ mode of such a cavity, which has a very high form factor of $\sim\,$0.81, as reported in\,\cite{Kuo:2019cps}. The TM$_{010}$ mode of the cylindrical design maintains azimuthal and longitudinal uniform radial field profiles, thus $C$ and $Q$ values close to those of the TM$_{010}$ mode of the aforementioned limit cavity are expected. Since the volume of such a cavity scales as $\lambda \cdot (r_{\rm in}+r_{\rm ext})/2$ instead of $\lambda^2$, it represents an extremely sensitive transducer for QCD axions especially in the high frequency region of the parameter space, where typically available solenoids are much wider than $\lambda$.

To use this cavity in a haloscope frequency tuning is necessary. Insertion of conductor or dielectric rods\,\cite{RevModPhys.75.777} inside the cavity quickly degrades the form factor of the mode, as do any geometrical perturbations that break the cylindrical symmetry of the cavity. The optimal solution is an uniform tuning of the gap width, similarly to what is done in thin shell cavities, but achieving this in a cylindrical geometry appears challenging. %as do any movements of one shell with respect of the other. %In such situations in particular, the field concentrates towards the wider part of the cavity and vacates the other parts where it is effectively in cutoff, leading to a decrease of the effective volume. To avoid this phenomenon and contemporarily obtain high frequency tuning, the optimal solution would be to preserve the cylindrical symmetry and only tune the radius of one of the cylinders, similarly to what is done in parallel plate thin shell cavities. While such a solution may be possible to accomplish it is undoubtedly challenging, especially in a cryogenic environment, so here we propose an alternative solution.
However, by substituting the cylinders with right prisms based on regular polygons with the same number of sides, the continuous azimuthal symmetry of the system becomes discrete. %, opening the path for a simple tuning based on relative rotation of the prisms. 
Thanks to this symmetry breaking, the azimuthal field distributions and resonant frequency of the transverse modes depend on the reciprocal angle between the two prisms. By simply rotating one prism with respect to the other, we can tune the cavity resonant frequency without excessively impacting the quality factor and the form factor of the mode.

For the present cavity design, the two nested coaxial prisms are enclosed within two flat endcaps. The cavity supports a (pseudo-)TM$_{010}$ axion-sensitive mode between the two prisms as depicted in Fig.\,\ref{fig:Cav_model_E_B_distr}, where electric and magnetic field distributions are shown.
As the cavity has constant section, the field distributions are longitudinally uniform. %The azimuthal profile is also rather uniform, as in the limit of large inner radius and infinite sides the cavity simplifies to an infinite rectangular box, with a very high form factor of $\sim$0.81, as reported in Ref.\,\cite{Kuo:2019cps}. % NB! Rivedere descrizione cavità: come è fatta? Partire di nuovo da shell circolari concentriche? Inoltre specificare che l'aspect ratio dev'essere >1  NB! for the TM110 mode of the rectangular cavity! Formalmente n on sembra lo stesso!

\subsection{\label{sec:tuning}Cavity Frequency Tuning}

\begin{figure*}[ht!]
     \centering
       \subcaptionbox{\small Electric field distribution, prisms with no reciprocal rotation.}{\includegraphics[width=.32\linewidth]{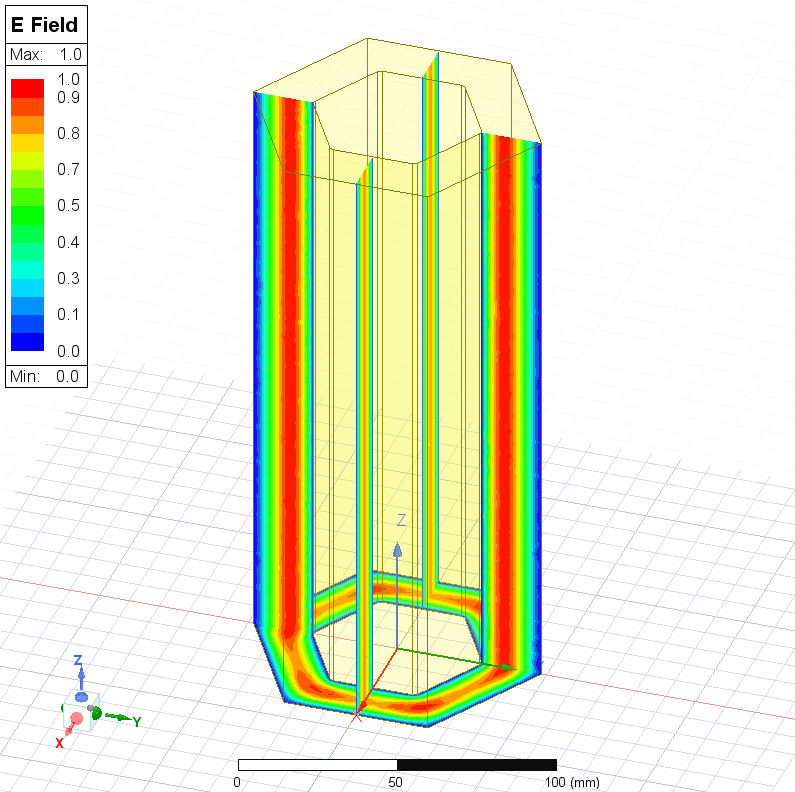}}\quad
       \subcaptionbox{\small Electric field distribution, prisms with 15 deg reciprocal rotation.}{\includegraphics[width=.32\linewidth]{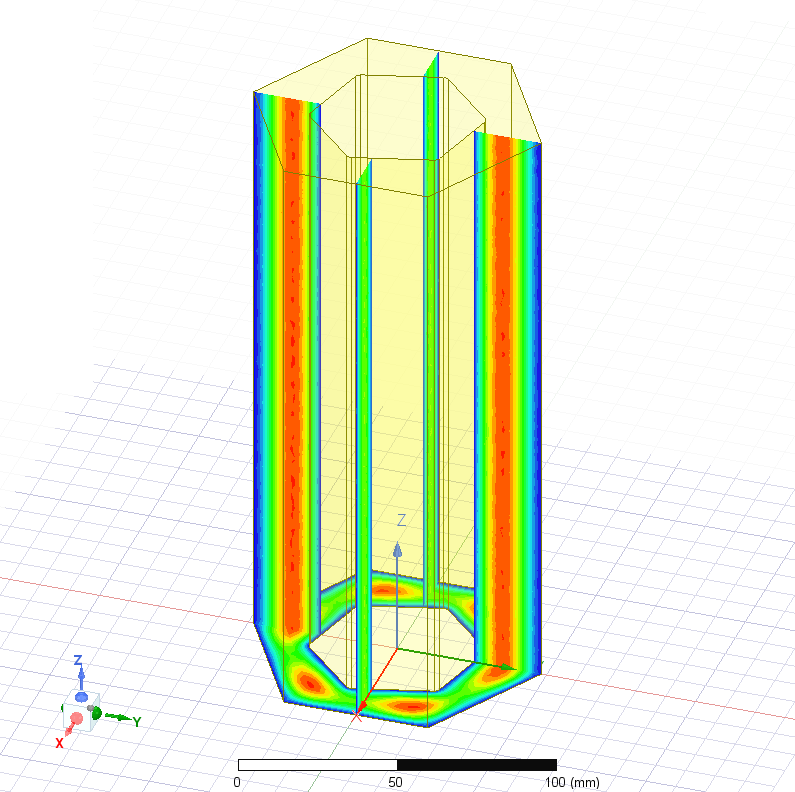}}\quad
       \subcaptionbox{\small Electric field distribution, prisms with 30 deg reciprocal rotation.}{\includegraphics[width=.32\linewidth]{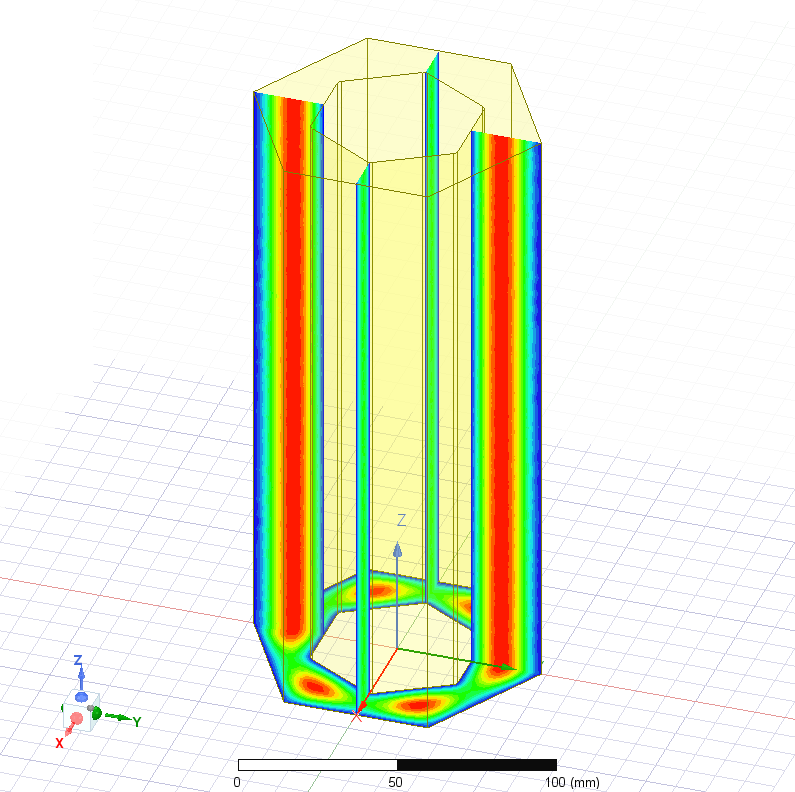}}\quad     
\caption{\small \label{fig:Cav_model_E_distr}Electric field profiles obtained from HFSS eigenmode simulation for the TM$_{010}$ mode. Cavity dimensions were $r_{\rm in}=39.14$\,mm and $r_{\rm ext}=55.92$\,mm for the internal and external circumradii and $L=210$\,mm for the length.}
\end{figure*}

%\begin{figure}[h]
%\centering
%\includegraphics[width=3.3in]{50mm_cavity_model.PNG}
%\caption{\small 50 mm-length toy model employed in comparing the cavity properties in function of the number of sides. The example with 6 sides is shown in figure.}
%\label{fig:50mm_cav_model}
%\end{figure}

%As previously stated, the key reason we are interested in using concentric polygons instead of the ideal case of two concentric circular conductors has to do with the proposed tuning. By promoting the circular prisms to polygonal prisms, and allowing one to rotate co-axially with respect to the other, we are able to (close-to) uniformly distribute the variation in the cavity radius that induces the tuning of the frequency of the $TM_{010}$ mode more uniformly, potentially allowing us to cover a wide frequency range with more than reasonable values of $C$ factor decrease. 

% From theoretical considerations using perturbation theory as a guide, we expect that 

%The length of the modeled prisms was 210\,mm, with chosen internal and external circumradiuses $r_{\rm in}=39.14$\,mm and $r_{\rm ext}=55.92$\,mm respectively. %
By using a finite element simulation (FEM) software package (Ansys HFSS) we study the properties of our resonator. In the cavity model we choose internal and external circumradiuses or $r_{\rm in}=39.14$\,mm and $r_{\rm ext}=55.92$\,mm respectively.
The gap width $d$ sets at first order the resonant frequency, while the number of sides determines the obtainable tuning range. % and by the ratio between the cavity inner and outer radius.
We define an angle $\phi$ to describe the relative rotation of the two prisms, setting $\phi=0$ when the prisms have parallel sides. Due to symmetry and cyclicity of the system, $\phi$ values span the range 0 to $\phi_{max}=\pi/N$.
Fig.\,\ref{fig:Cav_model_E_distr} shows the electric field profiles obtained from FEM simulations for different rotation angles. 
The rotation concentrates the electric field at the external prism vertexes, increasing the effective width of the cavity and in turn diminishing the cavity frequency.

\begin{figure}[h]
\centering
\includegraphics[height=3.4in,angle=-90]{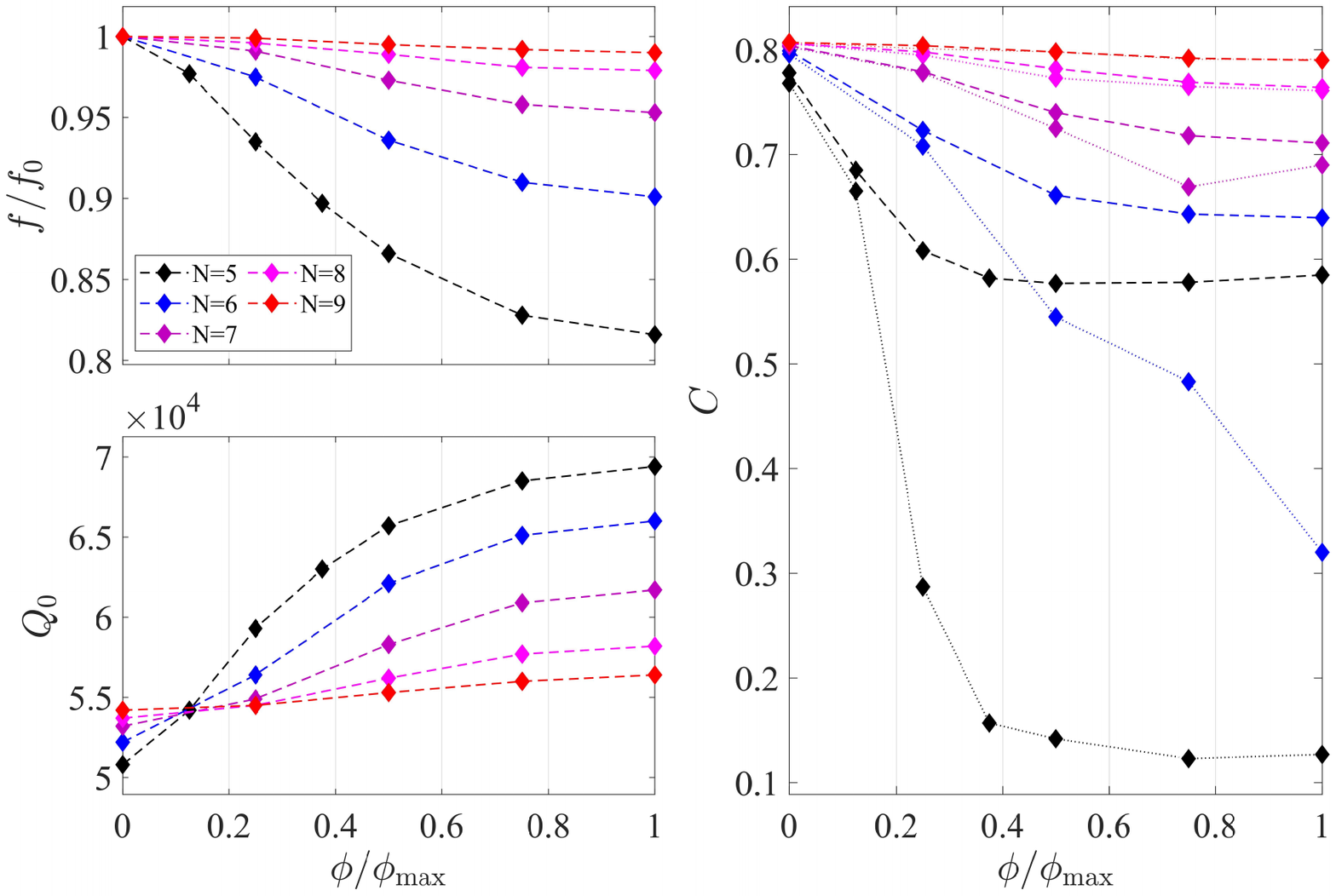} % Width is 3.33' if not rotated
\caption{\small Internal prism rotation and cavity parameters for different sides number. Left: normalized frequency shift and unloaded quality factor. Right: form factor. Results obtained with eigenmode simulations are plotted with dashed lines, while dotted lines indicate results obtained using driven modal simulations. Compared to the model shown in Fig.\,\ref{fig:Cav_model_E_distr}, we use a 50\,mm-long cavity to reduce computational costs.}
\label{fig:Cavity_parameters_vs_N_sides}
\end{figure}

In Fig.\,\ref{fig:Cavity_parameters_vs_N_sides} we plot the cavity parameters for polygonal resonators with $N=5$ to 9 sides, as a function of $\phi/\phi_{max}$.

%In order to be able to plot the results for cavities of different number of sides (and consequently, of different angular periodicity), the results are presented in function of the fractional angle of tuning $\Delta \phi$ calculated with respect to the maximum tuning angle. The starting position with $\Delta \phi$=0 corresponds to the prisms with parallel sides as in Fig.\,\ref{fig:NB! METTERE RIFERIMENTO CORRETTO}.

%A small number of sides would cause a too large symmetry breaking, resulting in reduced tuning as a consequence of a too quick degradation of the cavity parameters, while a too big number of sides causes the changes in effective radius to become too small, directly reducing the tuning range while leaving the cavity parameters essentially unperturbed.

%The used model for number of sides equal to 6 is represented in Fig.\,\ref{fig:50mm_cav_model}, where the prisms are in the position of maximum tuning. 
To reduce computational costs we used 50\,mm-length cavities with the same circumradii as in Fig.\,\ref{fig:Cav_model_E_distr}. %fixed $r_{\rm ext}=55.92$ mm and $r_{\rm in}=39.14$ mm. (NB! Da rivedere se non sono le stesse!) 
Despite the relatively short cavity length, the quality factor $Q_0$ is dominated by dissipation on the lateral faces. As the cavity wall gap changes with number of sides, the values of the resonant frequency $f_0$ of the TM$_{010}$ mode at $\phi=0$ are reported in Tab.\,\ref{tab:Fstart}.

\textbf{
\begin{table}[ht!]
\begin{center}
\caption{\label{tab:Fstart}\small Resonant frequency at $\phi=0$ for different sides numbers.}
\renewcommand{\arraystretch}{1.1} % Modify row spacing by a factor 1.1
\begin{tabular}{ |c|c|c|c|c|c| } 
 \hline
 Side number & 5 & 6 & 7 & 8 & 9 \\ 
 \hline
$f_0$ (GHz) & 10.820 & 10.165 & 9.798 & 9.574 & 9.425 \\ 
\hline
\end{tabular}
\end{center}
\end{table}}

%, with only the $Q$ factor being potentially impacted due to the different ratio of endcaps-to-sides dissipation. However, in this case the impact was nonetheless low, since dissipation on the lateral faces still dominates. % NB!!!!!! CONTROLLARE CHE SIA COSI' 

%\begin{center}
%\begin{tabular}{ |c|c|c|c|c|c|c| } 
% \hline
% Side number & $\Delta\phi$ & f (GHz) & $\Delta$f/$f_{0}$ & Q$_0$ & C$_{d.m.}$ & C$_{eig.}$ \\ 
% \hline
% \hline
%5 & 0 & 10.820 & 1 & 50800 & 0.768 & 0.778 \\ 
%\hline
%5 & 18 & 9.371 & 0.866 & 65700 & 0.142 & 0.577 \\ 
%\hline
%5 & 36 & 8.834 & 0.816 & 69400 & 0.127 & 0.585 \\ 
%\hline
%6 & 0 & 10.165 & 1 & 52200 & 0.796 & 0.799 \\ 
%\hline
%6 & 15 & 9.517 & 0.936 & 62100 & 0.545 & 0.661 \\ 
%\hline
%6 & 30 & 9.163 & 0.901 & 66000 & 0.320 & 0.6395 \\ 
%\hline
%7 & 0 & 9.798 & 1 & 53200 & 0.803 & 0.804 \\ 
%\hline
%7 & 12.86 & 9.531 & 0.973 & 58300 & 0.725 & 0.740 \\
%\hline
%7 & 25.71 & 9.333 & 0.953 & 61700 & 0.690 & 0.711 \\ 
%\hline
%8 & 0 & 9.574 & 1 & 53700 & 0.806 & 0.806 \\ 
%\hline
%8 & 11.25 & 9.464 & 0.989 & 56200 & 0.773 & 0.782 \\ 
%\hline
%8 & 22.5 & 9.369 & 0.979 & 58200 & 0.761 & 0.764 \\ 
%\hline
%9 & 0 & 9.425 & 1 & 54200 & 0.807 & 0.807 \\ 
%\hline
%9 & 10 & 9.378 & 0.995 & 55300 & 0.798 & 0.798 \\ 
%\hline
%9 & 20 & 9.335 & 0.990 & 56400 & 0.791 & 0.790 \\ 
%\hline
%\end{tabular}
%\end{center}

The change of the normalized frequency $f/f_0$ with $\phi$ is larger for smaller numbers of sides, as is the case for $Q_0$, which increases up to 30$\%$ at maximum tuning.
This is related to the concentration of the field towards the vertexes: at angles different from 0, the gap between external vertex and the face of the inner polygon increases for smaller sides number, while the opposite holds for the gap between the midpoint of the external faces and the inner vertexes. This in turn leads to increased mode confinement, greater frequency shifts and to smaller surfaces for the mode currents to dissipate on. %the mode is more confined near the vertexes of the external prism, leading to less dissipation. The trend is reversed at 0 tuning, this time due to the presence of sharper angles for the internal prism leasing to increased dissipation in the areas closer to its vertexes (while the dissipation in the other areas is mostly unchanged).

\begin{figure}[h]
\centering
\includegraphics[height=3.6in,angle=-90]{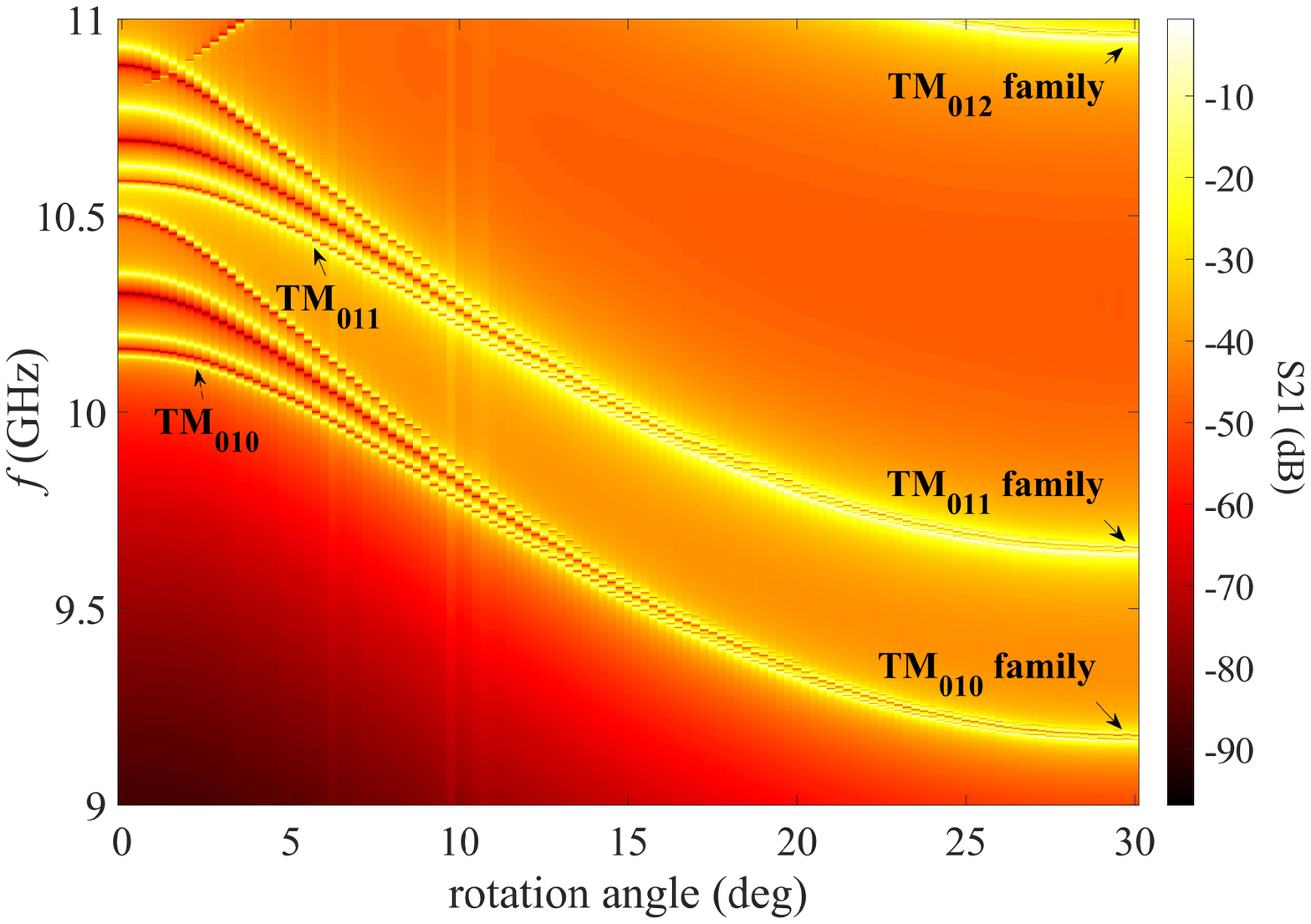}
\caption{\small Transmission S21 between two antennas on opposite endcaps as a function of tuning angle. A set of 121 driven modal simulations were performed setting copper conductivity at 4\,K as boundary condition. For the axion-sensitive TM$_{010}$ mode the maximum frequency shift is about 1\,GHz, with no mode mixings.}
\label{fig:S21_Heatmap}
\end{figure}

In the $C$ factor plot we also include the results of driven modal simulations, which calculate modal S-parameters and field profiles inside the cavity by driving the system through a lumped/wave port. % in terms of incident and reflected power of waveguide modes. % propagation through the cavity between two antennas takes place. 
Owing to electromagnetism's reciprocity, the driven modal approach allows for reliable $C$ factor calculation in multi-cell systems. In the present polygonal cavity, $C$ factor estimations via eigenmode simulation, although correct for the initial cavity configuration, could give incorrect results once the tuning angle is increased.
In fact, the polygonal resonator can be seen as a cyclic concatenation of identical cavities or cells, and the overall system behaviour depends on the coupling strength between adjacent cells. While in the case of strong coupling the system behaves as a single cavity, for couplings %close to or 
smaller than the mode linewidth the energy deposited in a cell will mostly be dissipated on its walls instead of being transmitted to the neighboring ones. %Let us now suppose $Q_e\gg Q_0$, where we denote with 1/Q$_e$ the coupling between adjacent cells so that for each of the N identical cells the loaded quality factor is $Q_L=(1/Q_0+2/Q_e)^{-1}$. In this limit the power transmitted from a cell at distance k to the cell containing the antenna is $\sum_{n=0}^{\infty} [(Q_L/Q_e)^{k+2n}+(Q_L/Q_e)^{N-k+2n}]\approx(Q_L/Q_e)^{min(k,k+N)}$%(Q$_L$/Q$_e$)$^k$+(Q$_L$/Q$_e$)$^{N-k}$, where we get two terms due to the circularity of the system. %More generally, this is the first order term of the series giving the correct result. Numerically approximated values can be computed for higher couplings, by means of tweaked tree diagrams or sviluppi di polinomi
As a consequence, for couplings close to or smaller than the mode linewidth, the eigenmode simulation yields a $C$ factor much higher than the real one. The latter effectively mimics the axion signal being equally transduced into each cell but fails to model the pickup of the signal from a single one%While in the case of strong coupling the system behaves as a single cavity, for couplings close to or smaller than the mode linewidth, a photon emitted in one of the cells farther from the antenna has increased probability to be absorbed by the cavity walls before reaching the antenna cell. The eigenmode simulation yields a $C$ factor much higher than the real one, since it effectively mimics the axion signal being equally transduced into each mode, but fails to model the pickup of the signal in a single cell. %NB! Provare a riscrivere come energia anzichè fotone: linewidth, energy released in one of the cells has increased absorption
\footnote{Nonetheless, provided each cell has its own tunable antenna, the present polygonal resonator becomes a cavity array in which the multiple cavities have $C\sim0.6$, $Q\sim60000$ and are all tuned together over a large range. In addition, for phase-matched readout, the cavity cells would only need an additional fine-tuning mechanism to compensate for small frequency differences between cells.}.% (i.e. a metallic rod inserted near the corner of the outer prism, so as to not interfere with the tuning of the cavity).
Mode crossings, i.e. regions where the axion-sensitive mode mixes with intruder modes, have been studied using driven modal simulations in the $N=6$ case. The resulting transmission spectra are shown in Fig.\,\ref{fig:S21_Heatmap}. 
%The effect of such a change is a wider spread of the longitudinal modes (in particular, considering that the length of the cavity is much larger than the width of the cavity, we can expect that an halving of the cavity roughly doubles the distance between the longitudinal modes closer to the fundamental one).
 The map shows two families of modes, each with four members that get closer at increasing tuning angles, which derive from the discretization of the TM$_{010}$ and TM$_{011}$ modes of the circular coaxial resonator. The lowest frequency modes in each family are those showing no zeros in the azimuthal direction. % stemming from the multicell geometry. %(In particular, for the hexagonal cavity these uneven configurations are the vertex-centered +++---, the vertex-centered +-+-+- and the side-centered +-+-+-.)
No mode crossings occur for the TM$_{010}$ mode, which is always well-coupled with the antennas. 
In addition, we note that the separation between the modes exceeds their linewidths even for high tuning angles.

\subsection{\label{sec:tolerances} Study of allowed tolerances}

%\begin{figure}[h]
%\centering
%\includegraphics[width=3.3in]{50mm_cavity_model.PNG}
%\caption{\small Caption}
%\label{fig:tolerance_study_cav_model}
%\end{figure}

\begin{figure}[h]
\centering
\includegraphics[width=3.75in,angle=00]{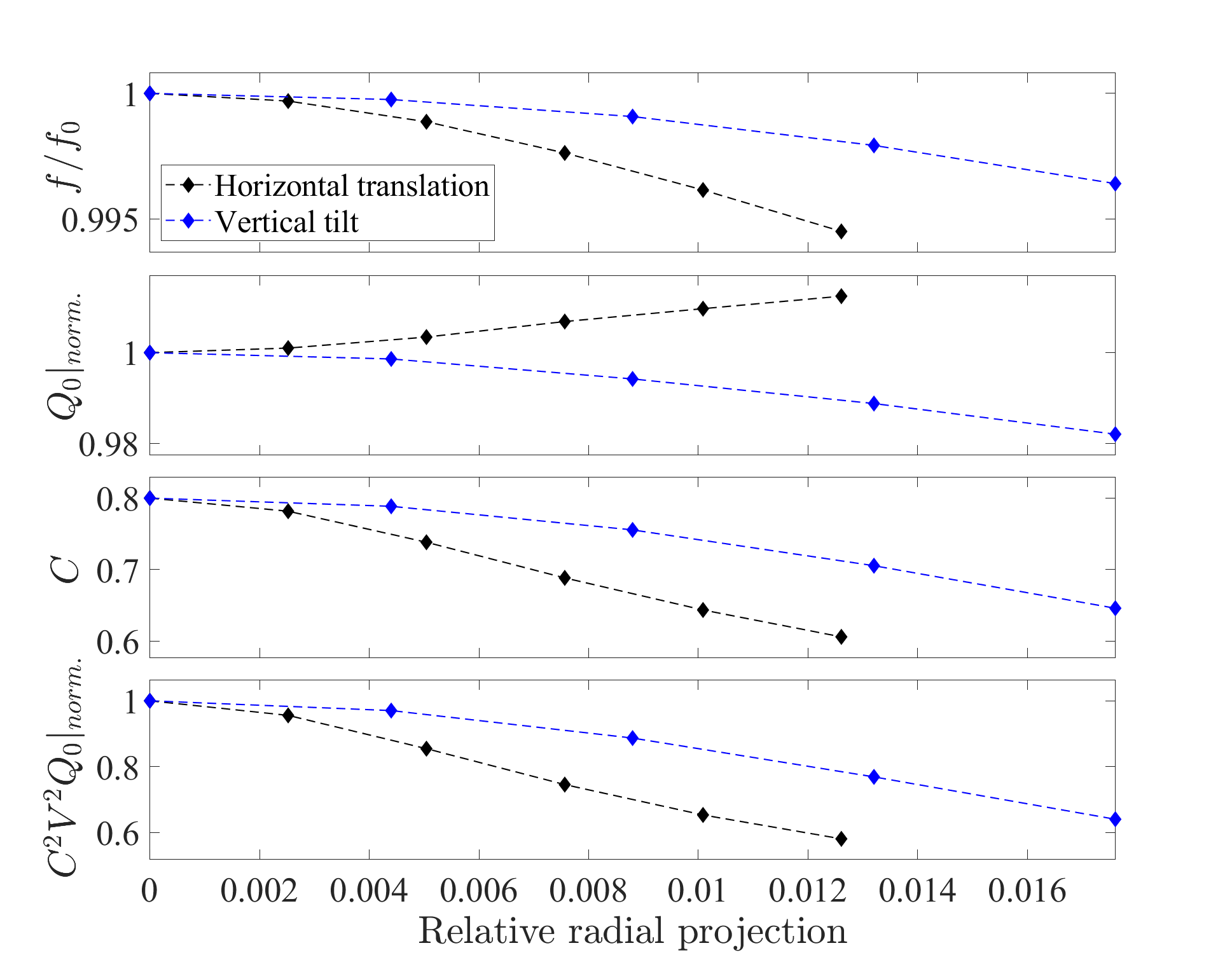} % Width is 3.33" if not rotated
\caption{\small \label{fig:Asymmetry_Tolerance_Study} Changes in the main cavity parameters due to varying degrees of reciprocal horizontal misalignment or tilting of the two prisms. In the case of the 10\,GHz cavity, the endpoint of the blue curve corresponds to a lateral shift of $\sim210\,\mu$m. From top to bottom: relative frequency shift, normalized quality factor $Q_0\vert_{norm.}$, $C$ factor and normalized figure of merit. $Q_0$ of the unperturbed cavity was 62000.}% The tuning angle of the cavity was 0 deg. The employed cavity had external circumradius $r_{\rm ext}$=48.3 mm, internal circumradius $r_{\rm int}$=28.47 mm and cavity height $h_{cav}$=200 mm, resulting in a base resonant frequency for the TM$_{010}$ mode of 8.529 GHz, $Q_0$ of 62300 and C of 0.7999.}
\end{figure}

%\begin{figure*}[ht!]
%     \centering
%       \begin{tabular}{@{}c@{}}
%    \includegraphics[width=.33\linewidth,angle=-90]{Asymmetry_Tolerance_Study_f.pdf} \\[\abovecaptionskip]
%    \small A. Relative frequency shift.
%  \end{tabular}
%  \begin{tabular}{@{}c@{}}
%    \includegraphics[width=.33\linewidth,angle=-90]{Asymmetry_Tolerance_Study_Q0.pdf} \\[\abovecaptionskip]
%    \small B. Normalized quality factor.
%  \end{tabular}
%  \begin{tabular}{@{}c@{}}
%    \includegraphics[width=.33\linewidth,angle=-90]{Asymmetry_Tolerance_Study_C.pdf} \\[\abovecaptionskip]
%    \small C. $C$ factor.
%  \end{tabular}
%  \begin{tabular}{@{}c@{}}
%    \includegraphics[width=.33\linewidth,angle=-90]{Asymmetry_Tolerance_Study_F_merit.pdf} \\[\abovecaptionskip]
%    \small D. Normalized figure of merit.
%  \end{tabular}
%\caption{\small \label{fig:Asymmetry_Tolerance_Study} Changes in the main cavity parameters due to varying degrees of reciprocal horizontal misalignment or tilting of the two prisms. The endpoint of the blue curve corresponds to a lateral shift of 250 micron.}
%\end{figure*}

Complex cavity designs are more likely susceptible to deviations from the ideal geometry. We evaluate the required mechanical tolerances needed in this geometry by considering horizontal misalignment and tilts between the two prisms at $\phi=0$. 
%The employed cavity model was slightly modified by increasing the length of the inner conductor so that it could be tilted without geometry issues; 
Fig.\,\ref{fig:Asymmetry_Tolerance_Study} reports the results for cavity parameters and figure of merit. We move the axis of the inner prism in the plane defined by the axis of the second prism and one of its vertexes. We separately study the effect of horizontal misalignment and of tilts about the center of the cavity.
%We study the effect of misalignment by displacing the inner prism axis along the line through two horizontally-adjacent vertexes. Tilts were separately studied by rotating the inner prism about the normal of one of its lateral faces passing through the prism center.

The two datasets are presented in the same plot where the horizontal axis is the relative radial projection, i.e. the displacement of the intercept of the inner prism axis with the top endcap, normalized to the prisms circumradius difference. %In this representation, the maximum value of the abscissa corresponds to a shift of 250\,$\mu$m. In fact in a real model, the inner prism will have some tolerance around its axis to allow for the rotational tuning of the cavity. If we hypothesize now that this tolerance will be isotropic around the axis and equal at both endcaps, and that the inner prism will at worst either rest touching the same side of the gap on both endcaps, or rest touching one side on one endcap, and the opposite one on the other, we end up with a situation similar to the one represented here (with the minor oversight that the prism would rest in the middle of the sides, not on a vertex).

Horizontal misalignments significantly impact the $C$ factor and thus are the main source of cavity figure of merit degradation. Quality factor and resonant frequency changes are much smaller.
%The strongest degradation in the performance of the cavity comes from a misalignment of the prisms, with the changes in the width of the cavity that rapidly send part of the cavity in the evanescent field regime, with a consequent sharp decrease of the form factor. The mechanism for the $C$ factor decrease is the same in the case of a tilt, but the onset is delayed due to the reduced asymmetry of the system (at the center, the cavity keeps the same width as before, with the asymmetry in the cavity width only being maximum at the two endcaps instead that all along the cavity). The changes in the quality factor and in the resonant frequency are comparably negligible, the latter being not surprising considering the exponential nature of the evanescent cutoff phenomenon.
To limit the scan rate degradation due to horizontal translations within 10$\%$, the minimum tolerance needed would be of about 75 $\mu$m, within the capabilities of computerized numerical control machines.

%Finally, it is also important to note that the particular choice of the movement situations does not at all exhaust the complexity of nonidealities that can happen in a real model, where we will have, in general, a combination of both effects, with translations and rotations happening along arbitrary lines/planes and the two prisms likely rotated with respect to one each other by a nonzero azimuthal tuning angle. Nonetheless, we think that the two reported datasets are sufficient to have an idea of the scale of the expected effects (especially considering we are working with a closed and very simple geometrically perfect model which is itself by definition not buildable). Some combinations between tilts and translations have also been tested, but were not reported for graphical representation difficulties; the obtained results were consistent with a linear combination of translations and rotations.

%Here we need to start with simulations so we are more quantitative.
%
%-> we discuss the Q
%
%-> real C vs number of sides at set outer/inner radius ratio
%
%-> also discussion of the effect of different $Q$ factor of the single cavities on the "multicavity" with cavities at fixed coupling (i.e., fixed geometry)
%
%-> Delta f at set delta $C$ for different tuning mechanisms
%
%-> study of cavity nonidealities: effect of misalignments and tilts

\section{\label{sec:measurements}Experimental results}
\begin{figure}[h]
\centering
       {\includegraphics[width=.5127\linewidth]{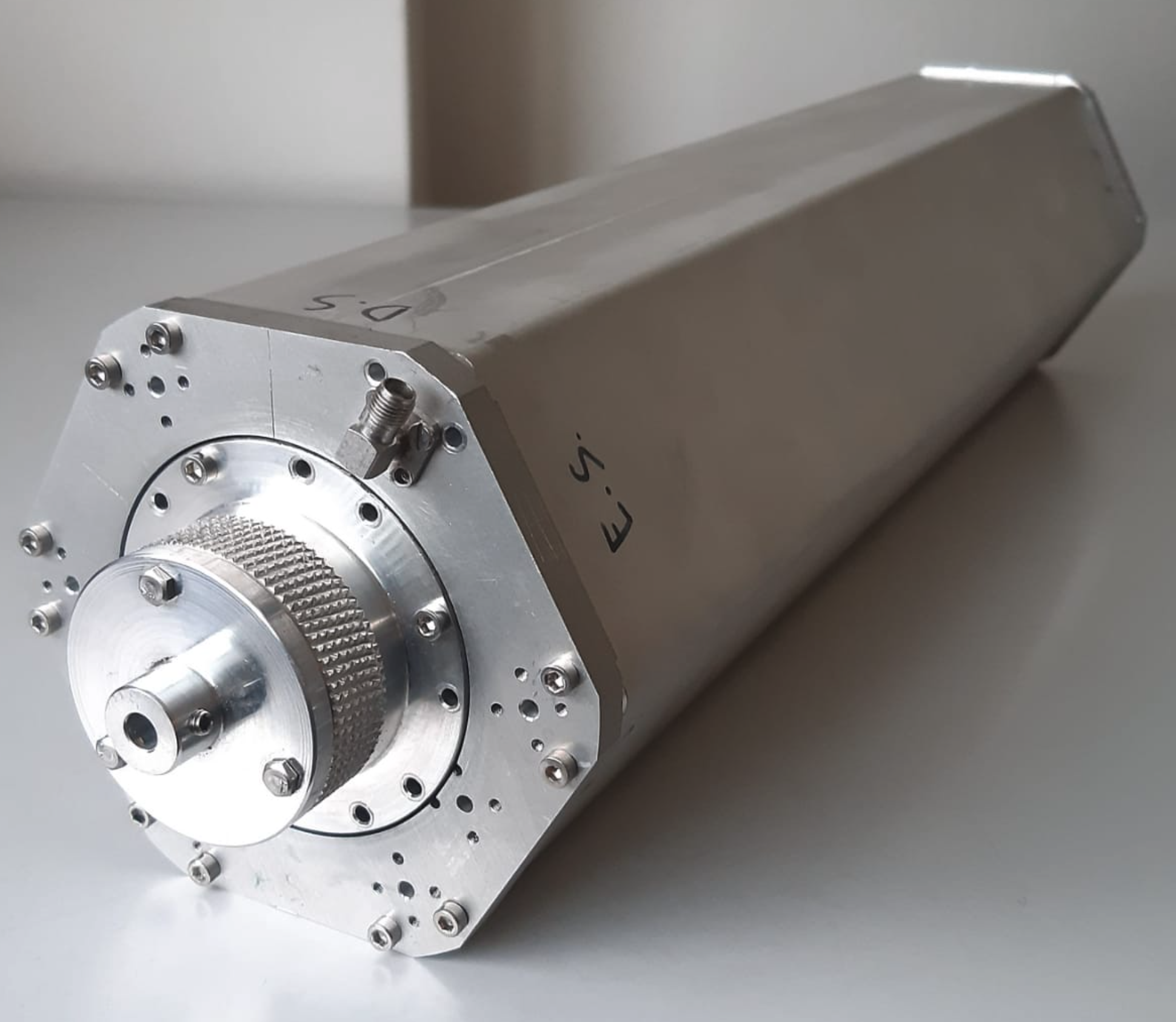}}\quad
       {\includegraphics[width=.4493\linewidth]{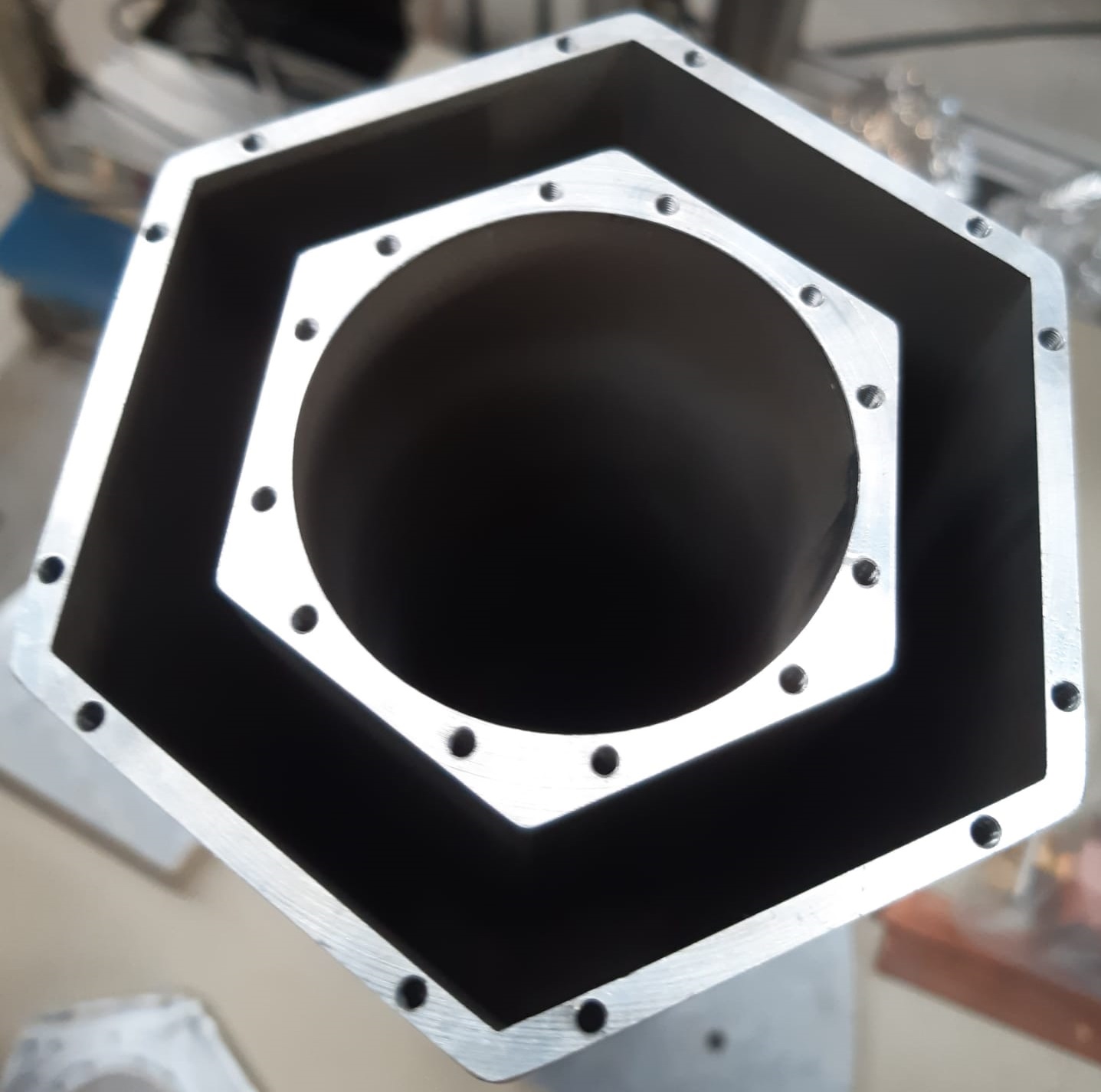}}\quad
\caption{\small The prototype hexagonal coaxial resonator. Left: external view. Right: internal view.}% after coating with copper tape. }
\label{fig:cavity_prototype}
\end{figure}

A full-scale prototype of the cavity was obtained from bulk aluminium metal via single-pass wire electrodischarge machining (wire EDM) and is shown in Fig.\,\ref{fig:cavity_prototype}. The 420\,mm-long cavity has $r_{\rm in}$ and $r_{\rm ext}$ of 55.91\,mm and 39.14\,mm, respectively. The longitudinal edges of the inner and outer prism are filleted with radii of 2\,mm and 3\,mm, respectively. In this prototype, the endcaps are divided into two pieces through a 50\,$\mu$m-wide circular gap, allowing for rotation of the inner prism via a knob. Each endcap hosts 8 apertures used for antenna insertion and dielectric bead pull characterization\,\cite{som2011bead}. Of those, 6 were positioned in correspondence of the outer prism vertexes, while the other two were in the middle of a side. All apertures were radially positioned to lay at the center of the cavity gap.  %Part of the employed setup for the bead pulling measurement is also visible in the background, with the XXX mm-long, YYY\,mm-thick polyethylene cylindrical bead visible on top. NB! scrivere dimensioni e materiale della bead!!!!

% NB! Scrivere figura (a), (b), (c) nella fig.7; 

\begin{figure}[h]
\centering
\includegraphics[height=3.5in,angle=-90]{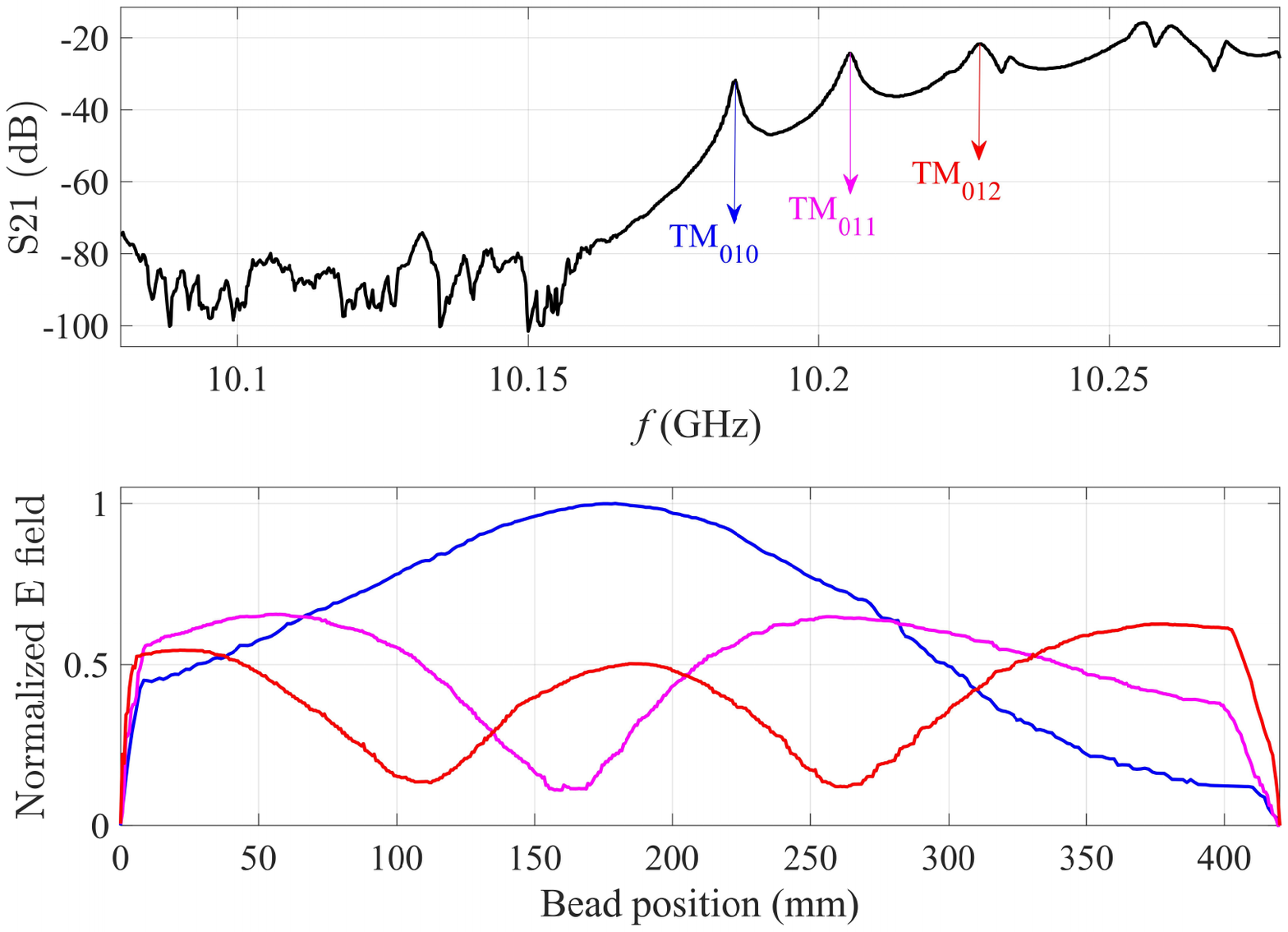}
\caption{\small Experimental results. The upper panel reports the transmission spectrum measured at $\phi=0$, where we also indicate the position of the first three TM$_{01n}$ modes. The corresponding measured electric field profiles are shown in the lower panel with the same colours.}
%\caption{\small Experimental setup and results. The employed cavity is shown in box (a), with the support employed for the bead pull measurement visible in the background. Box (b) reports the transmission spectrum measured at $\phi=0$, where we also indicate the position of the first three TM$_{01n}$ modes. The corresponding measured electric field profiles are shown in box (c).}
\label{fig:experimental_results}
\end{figure}

The measured transmission spectrum of the cavity at $\phi=0$ is shown in Fig.\,\ref{fig:experimental_results}. The TM$_{010}$ and TM$_{011}$ modes are well-isolated and the spectrum is clean before TM$_{010}$, in agreement with simulation results shown in Fig.\,\ref{fig:Asymmetry_Tolerance_Study}.% The third mode is instead perturbed by a couple of intruder modes, even though bead pull results (see Fig.\,\ref{fig:experimental_results}\,(c)) show no significant coupling to them.

The frequency of the TM$_{010}$ mode measured at room temperature was 10.1855\,GHz, to be compared with 10.14\,GHz yielded from numerical simulations. The measured quality factor was $\sim4200$, with negligible couplings on both antennas. In the $\phi=0$ configuration, the field profiles and transmission spectra were independent of antenna positioning.
The full tuning range for a rotation from 0 to $\phi_{max}$ was about 1\,GHz, as expected from simulations.% with the first mode remaining clean for most of the range.

Measurements of field profiles evidenced the presence of geometrical nonidealities.
Fig.\,\ref{fig:experimental_results} shows bead pull results for the main mode and the two most adjacent ones. The TM$_{010}$ mode longitudinal field profile peaks before the cavity center, and its magnitude at the endcaps differs significantly. Marked asymmetries are also visible in the TM$_{011}$ and TM$_{012}$ modes. The cavity prototype is therefore longitudinally asymmetric. Concerning the azimuthal field distribution, we monitored the maximum field amplitudes for the TM$_{010}$ mode, evidencing variations up to a factor 3.
Numerical control machine measurements revealed deviations from design dimensions of up to a maximum of 0.14\,mm. Minor geometry adjustments in the prism alignment resulted in a small frequency variation and no significant changes in the quality factor.

%Due to the discovered nonidealities, an effort was made to fix the cavity shape with further machining. After this recentering step, the cavity was remounted and internally coated with copper tape strips. The copper coated cavity showed a room temperature unloaded $Q$ factor of 9400, not dissimilar from the simulated value of QQQQQ. After cooling to 77 K, we observed a $Q$ factor of 17800, again not dissimilarly from the expected value of QQQQQ (NB! $Q$ da SIMULAZIONE CAVITA' COMPLETA)

%The prisms centering in the prototype was thus improved, 

Subsequently the cavity was internally coated with copper tape strips, which improved the room temperature $Q_0$ to 9400, to be compared with the value of 10900 obtained for a bulk OHFC copper cavity from simulations. Cooling to 77\,K improved $Q_0$ to\,17800, smaller than the simulated value of 32900. Among the possible reasons for this discrepancy may be a lower than expected improvement in RF conductivity for the copper tape, or additional geometrical nonidealities in the prototype induced by thermal contractions.  %further geometric nonidealities caused by the movement degree of freedom allowed to the two prisms.% or to a greater impact of the losses in the gaps present in the endcaps. The latter hypothesis however is strongly disfavoured, since the improvement in conductivity of aluminium should be bigger than the one of copper, leading to it representing a smaller fraction of the dissipations.

\section{\label{sec:comparisons}Cavity comparisons}

\begin{figure}[ht!]
\includegraphics[height=3.3in,angle=-90]{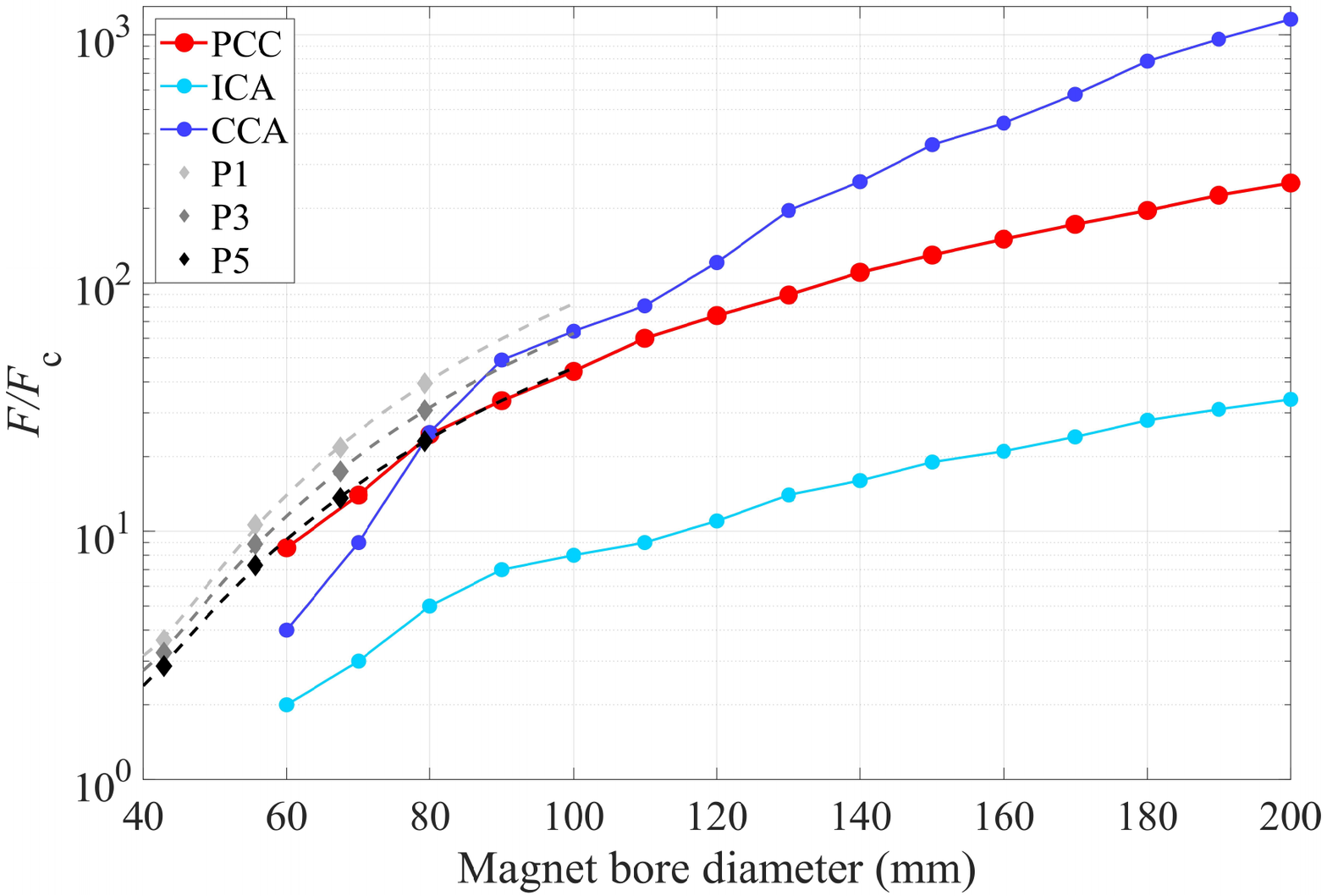}
\caption{\small \raggedright Comparison between normalized figures of merit for different resonant cavity systems filling a magnet bore in a 10\,GHz haloscope. PCC - polygonal coaxial cavity; ICA - independent cavity array; CCA - coherent cavity array. P1, P3 and P5 - pizza cavities employing diaphragms of 1, 3 and 5 mm thickness, respectively. The corresponding dashed lines are second order polynomial fits, while the solid lines only represent a guide to the eye.}
\label{fig:cavity_comparison}
\end{figure}

In Fig.\,\ref{fig:cavity_comparison} we report the overall figure of merit $F$ (see eq.\,\ref{eq:fig_of_merit}) of the polygonal coaxial cavity for various magnet bore values. The working frequency has been set to 10\,GHz and $F$ is normalized to the value $F_c$ obtained for a single cylindrical cavity of radius 11.48\,mm resonating at the same frequency. For comparison, we also add the performances of some cavity setups described in the introduction: pizza cavities and cavity arrays. In the latter case, the normalized factor of merit is $K$ and $K^2$ in the case of independent and coherent cavity arrays, respectively.
The following assumptions have been made:
\begin{itemize}
\item the external wall thickness is 3 mm for all cavities;
\item to fix TM$_{010}$ at 10\,GHz for the polygonal cavity, the gap thickness is kept constant for increasing magnet bore, while the number of sides is qualitatively scaled as the square root of the magnet bore. The external radius is adjusted to the maximum possible value. The $C$ factor is 0.8 and we use a conservative value of 50000 for the quality factor; 
\item the data series for the pizza cavities are based on published data\,\cite{Jeong:2017hqs} obtained with 2, 4, 6 and 8 sections, rescaling the cavity dimensions to ensure resonance at 10\,GHz. Their quality factors are respectively 68000, 62000, 55000, 50000 with $C=0.65$; %NB! Q delle cavità? dove diciamo che assumiamo tenda a 50000?
\item for the cavity arrays, we set cylindrical geometry with $Q_0=100000$ and C=0.69. The cavities are arranged to optimally fill the available volume. Arrays of 2 to 34 cavities can be hosted within the bore.
\end{itemize}

% NB! Aggiungere le freccette per disegnetti con i numeri di cavità
%NB! Togliere dashed line per ICA?, maybe blu e rosso con sfumature più chiare?

In Fig.\,\ref{fig:cavity_comparison}, we identify three regions of interest for high frequency haloscopes. 
Below 80\,mm-diameter magnet bore, the pizza cavity is the best approach, as arrays of cylindrical cavities have low tessellation efficiency. The performance of the present polygonal coaxial cavity is comparable to a pizza cavity using 5\,mm-thick diaphragms (P5 in Fig.\,\ref{fig:cavity_comparison}). % but as pizza cavities with such numbers of subdivisions have already been tested, there is no reason to prefer it in this range.
Between 80 and 120\,mm the polygonal cavity approach is a competitive solution for its higher simplicity compared to pizza cavities with 10 to 16 subdivisions and arrays of 5 to 11 cavities.  
In principle, arrays of cylindrical cavities would give the best figure of merit for bore magnets greater than $\sim120\,$mm. However, one should be able to coherently readout a number of cavities ranging between 11 and 34, while the polygonal coaxial resonator is a practical solution.
Moreover, multiple cavity arrays can be based also on nested polygonal coaxial cavities, greatly reducing the total number of cavities to be operated in parallel.

%NB! Dopo ricordiamo di dire qualcosa di questo genere!
%In conclusion, the polygonal cavity allows to reach $C$ factors close to 0.8 at any arbitrary frequency and with an arbitrary inner radius, allowing to
%provided the symmetry breaking introduced by the presence of a finite number of sides not be too extreme,

\section{\label{sec:conclusions} Conclusions}

We introduced the tunable polygonal coaxial axion haloscope, and discussed it as a practical solution to the issue of the poor scaling of the scan rate with volume, particularly challenging in haloscopes in the range above $\sim 5$ GHz. Compared to arrays of cavities, and the pizza cavity concept, this resonator presents a significant figure of merit and allows for a practical and effective tuning method by relative rotation of the internal prism with up to 5\% tuning depth. 

FEM simulations were employed to study relevant cavity parameters over the tuning range. No mode mixings were observed. We also estimated the required mechanical tolerances, which were found to be within the capabilities of commonly available numerical control machines. %ti hanno permesso di capire che tolleranze servono e bla bla bla 

A prototype aluminium hexagonal cavity was built and tested. The axion-sensitive mode was found near the expected frequency and was isolated for the full expected tuning range. Bead pull measurements were used to map the longitudinal field profiles and the azimuthal field distribution inside the resonator, showing unwanted asymmetries due to geometrical nonidealities of the prototype. More precise machining techniques, such as numerical control, will therefore be needed for future realizations. The cavity's inner surfaces were covered with copper tape strips to increase its quality factor. At room temperature, this allowed to reach $Q_0\sim10^4$, not far from the expected value. At liquid nitrogen temperature, the measured $Q_0$ of the coated cavity was short of the simulated value for reasons not yet completely understood. Most likely, this is due to a limited RF low temperature conductivity of the copper strips, or to variation of RF contact in the endcaps due to thermal contractions. The tuning mechanism should therefore be designed for cryogenic measurements, for instance including ball bearing guides. Flexible gaskets or quarter-lambda chokes should also be considered to ensure good RF connections. %NB! Ripetere che le nonidealità geometria erano dovute alla tecnica di machining del pezzo

In conclusion, the polygonal resonator represents an effective 3D resonator for axion dark matter haloscope searches at high frequency.
%Perspectives and potential uses of the cavity
%As a closing note, we briefly touch on the future perspectives and potential uses of the presented cavity. 
Much higher quality factors up to the range of millions could be reached by coating the internal walls with high temperature superconductor strips (see\,\cite{PhysRevApplied.17.L061005}).
Additionally, even more efficient magnet bore occupation can be obtained by nested multiple polygonal coaxial cavities.

\begin{acknowledgments}
This material was based upon work supported by INFN (QUAX experiment).

We are grateful to E. Berto (University of Padova and INFN) who substantially contributed to the mechanical realization of this cavity. A. Benato, A. Pitacco and M. Rebeschini of INFN Padova did part of the mechanical work and M. Zago (INFN) the mechanical drawings. The contribution of F. Calaon and M. Tessaro (INFN) to the measurement set-up is gratefully acknowledged. The authors also thank A. Palmieri for useful discussions on the cavity concepts and M. Comunian for providing the server for FEM analysis.

\end{acknowledgments}

%\nocite{*}
%\bibliography{TM010_Coaxial_Polygonal_Cavity}% Produces the bibliography via BibTeX.
\bibliography{pseudo-TM010_Coaxial_Polygonal_Cavity}% Produces the bibliography via BibTeX.

\providecommand{\noopsort}[1]{}\providecommand{\singleletter}[1]{#1}%
\begin{thebibliography}{39}%
\makeatletter
\providecommand \@ifxundefined [1]{%
 \@ifx{#1\undefined}
}%
\providecommand \@ifnum [1]{%
 \ifnum #1\expandafter \@firstoftwo
 \else \expandafter \@secondoftwo
 \fi
}%
\providecommand \@ifx [1]{%
 \ifx #1\expandafter \@firstoftwo
 \else \expandafter \@secondoftwo
 \fi
}%
\providecommand \natexlab [1]{#1}%
\providecommand \enquote  [1]{``#1''}%
\providecommand \bibnamefont  [1]{#1}%
\providecommand \bibfnamefont [1]{#1}%
\providecommand \citenamefont [1]{#1}%
\providecommand \href@noop [0]{\@secondoftwo}%
\providecommand \href [0]{\begingroup \@sanitize@url \@href}%
\providecommand \@href[1]{\@@startlink{#1}\@@href}%
\providecommand \@@href[1]{\endgroup#1\@@endlink}%
\providecommand \@sanitize@url [0]{\catcode `\\12\catcode `\$12\catcode `\&12\catcode `\#12\catcode `\^12\catcode `\_12\catcode `\%12\relax}%
\providecommand \@@startlink[1]{}%
\providecommand \@@endlink[0]{}%
\providecommand \url  [0]{\begingroup\@sanitize@url \@url }%
\providecommand \@url [1]{\endgroup\@href {#1}{\urlprefix }}%
\providecommand \urlprefix  [0]{URL }%
\providecommand \Eprint [0]{\href }%
\providecommand \doibase [0]{https://doi.org/}%
\providecommand \selectlanguage [0]{\@gobble}%
\providecommand \bibinfo  [0]{\@secondoftwo}%
\providecommand \bibfield  [0]{\@secondoftwo}%
\providecommand \translation [1]{[#1]}%
\providecommand \BibitemOpen [0]{}%
\providecommand \bibitemStop [0]{}%
\providecommand \bibitemNoStop [0]{.\EOS\space}%
\providecommand \EOS [0]{\spacefactor3000\relax}%
\providecommand \BibitemShut  [1]{\csname bibitem#1\endcsname}%
\let\auto@bib@innerbib\@empty
%</preamble>
\bibitem [{\citenamefont {Weinberg}(1978)}]{PhysRevLett.40.223}%
  \BibitemOpen
  \bibfield  {author} {\bibinfo {author} {\bibfnamefont {S.}~\bibnamefont {Weinberg}},\ }\bibfield  {title} {\bibinfo {title} {A new light boson?},\ }\href {https://doi.org/10.1103/PhysRevLett.40.223} {\bibfield  {journal} {\bibinfo  {journal} {Phys. Rev. Lett.}\ }\textbf {\bibinfo {volume} {40}},\ \bibinfo {pages} {223} (\bibinfo {year} {1978})}\BibitemShut {NoStop}%
\bibitem [{\citenamefont {Wilczek}(1978)}]{PhysRevLett.40.279}%
  \BibitemOpen
  \bibfield  {author} {\bibinfo {author} {\bibfnamefont {F.}~\bibnamefont {Wilczek}},\ }\bibfield  {title} {\bibinfo {title} {Problem of strong $p$ and $t$ invariance in the presence of instantons},\ }\href {https://doi.org/10.1103/PhysRevLett.40.279} {\bibfield  {journal} {\bibinfo  {journal} {Phys. Rev. Lett.}\ }\textbf {\bibinfo {volume} {40}},\ \bibinfo {pages} {279} (\bibinfo {year} {1978})}\BibitemShut {NoStop}%
\bibitem [{\citenamefont {Peccei}\ and\ \citenamefont {Quinn}(1977{\natexlab{a}})}]{PhysRevLett.38.1440}%
  \BibitemOpen
  \bibfield  {author} {\bibinfo {author} {\bibfnamefont {R.~D.}\ \bibnamefont {Peccei}}\ and\ \bibinfo {author} {\bibfnamefont {H.~R.}\ \bibnamefont {Quinn}},\ }\bibfield  {title} {\bibinfo {title} {$\mathrm{CP}$ conservation in the presence of pseudoparticles},\ }\href {https://doi.org/10.1103/PhysRevLett.38.1440} {\bibfield  {journal} {\bibinfo  {journal} {Phys. Rev. Lett.}\ }\textbf {\bibinfo {volume} {38}},\ \bibinfo {pages} {1440} (\bibinfo {year} {1977}{\natexlab{a}})}\BibitemShut {NoStop}%
\bibitem [{\citenamefont {Peccei}\ and\ \citenamefont {Quinn}(1977{\natexlab{b}})}]{PhysRevD.16.1791}%
  \BibitemOpen
  \bibfield  {author} {\bibinfo {author} {\bibfnamefont {R.~D.}\ \bibnamefont {Peccei}}\ and\ \bibinfo {author} {\bibfnamefont {H.~R.}\ \bibnamefont {Quinn}},\ }\bibfield  {title} {\bibinfo {title} {Constraints imposed by $\mathrm{CP}$ conservation in the presence of pseudoparticles},\ }\href {https://doi.org/10.1103/PhysRevD.16.1791} {\bibfield  {journal} {\bibinfo  {journal} {Phys. Rev. D}\ }\textbf {\bibinfo {volume} {16}},\ \bibinfo {pages} {1791} (\bibinfo {year} {1977}{\natexlab{b}})}\BibitemShut {NoStop}%
\bibitem [{\citenamefont {Irastorza}\ and\ \citenamefont {Redondo}(2018)}]{IRASTORZA201889}%
  \BibitemOpen
  \bibfield  {author} {\bibinfo {author} {\bibfnamefont {I.~G.}\ \bibnamefont {Irastorza}}\ and\ \bibinfo {author} {\bibfnamefont {J.}~\bibnamefont {Redondo}},\ }\bibfield  {title} {\bibinfo {title} {New experimental approaches in the search for axion-like particles},\ }\href {https://doi.org/https://doi.org/10.1016/j.ppnp.2018.05.003} {\bibfield  {journal} {\bibinfo  {journal} {Progress in Particle and Nuclear Physics}\ }\textbf {\bibinfo {volume} {102}},\ \bibinfo {pages} {89} (\bibinfo {year} {2018})}\BibitemShut {NoStop}%
\bibitem [{\citenamefont {Workman}\ and\ \citenamefont {Others}(2022)}]{Workman:2022ynf}%
  \BibitemOpen
  \bibfield  {author} {\bibinfo {author} {\bibfnamefont {R.~L.}\ \bibnamefont {Workman}}\ and\ \bibinfo {author} {\bibnamefont {Others}} (\bibinfo {collaboration} {Particle Data Group}),\ }\bibfield  {title} {\bibinfo {title} {{Review of Particle Physics}},\ }\href {https://doi.org/10.1093/ptep/ptac097} {\bibfield  {journal} {\bibinfo  {journal} {PTEP}\ }\textbf {\bibinfo {volume} {2022}},\ \bibinfo {pages} {083C01} (\bibinfo {year} {2022})}\BibitemShut {NoStop}%
\bibitem [{\citenamefont {Buschmann}\ \emph {et~al.}(2022)\citenamefont {Buschmann}, \citenamefont {Foster}, \citenamefont {Hook}, \citenamefont {Peterson}, \citenamefont {Willcox}, \citenamefont {Zhang},\ and\ \citenamefont {Safdi}}]{Buschmann:2021sdq}%
  \BibitemOpen
  \bibfield  {author} {\bibinfo {author} {\bibfnamefont {M.}~\bibnamefont {Buschmann}}, \bibinfo {author} {\bibfnamefont {J.~W.}\ \bibnamefont {Foster}}, \bibinfo {author} {\bibfnamefont {A.}~\bibnamefont {Hook}}, \bibinfo {author} {\bibfnamefont {A.}~\bibnamefont {Peterson}}, \bibinfo {author} {\bibfnamefont {D.~E.}\ \bibnamefont {Willcox}}, \bibinfo {author} {\bibfnamefont {W.}~\bibnamefont {Zhang}},\ and\ \bibinfo {author} {\bibfnamefont {B.~R.}\ \bibnamefont {Safdi}},\ }\bibfield  {title} {\bibinfo {title} {{Dark matter from axion strings with adaptive mesh refinement}},\ }\href {https://doi.org/10.1038/s41467-022-28669-y} {\bibfield  {journal} {\bibinfo  {journal} {Nature Commun.}\ }\textbf {\bibinfo {volume} {13}},\ \bibinfo {pages} {1049} (\bibinfo {year} {2022})},\ \Eprint {https://arxiv.org/abs/2108.05368} {arXiv:2108.05368 [hep-ph]} \BibitemShut {NoStop}%
\bibitem [{\citenamefont {Sikivie}(1983)}]{PhysRevLett.51.1415}%
  \BibitemOpen
  \bibfield  {author} {\bibinfo {author} {\bibfnamefont {P.}~\bibnamefont {Sikivie}},\ }\bibfield  {title} {\bibinfo {title} {Experimental tests of the "invisible" axion},\ }\href {https://doi.org/10.1103/PhysRevLett.51.1415} {\bibfield  {journal} {\bibinfo  {journal} {Phys. Rev. Lett.}\ }\textbf {\bibinfo {volume} {51}},\ \bibinfo {pages} {1415} (\bibinfo {year} {1983})}\BibitemShut {NoStop}%
\bibitem [{\citenamefont {Kim}\ \emph {et~al.}(2020{\natexlab{a}})\citenamefont {Kim}, \citenamefont {Jeong}, \citenamefont {Youn}, \citenamefont {Kim},\ and\ \citenamefont {Semertzidis}}]{Kim:2020kfo}%
  \BibitemOpen
  \bibfield  {author} {\bibinfo {author} {\bibfnamefont {D.}~\bibnamefont {Kim}}, \bibinfo {author} {\bibfnamefont {J.}~\bibnamefont {Jeong}}, \bibinfo {author} {\bibfnamefont {S.}~\bibnamefont {Youn}}, \bibinfo {author} {\bibfnamefont {Y.}~\bibnamefont {Kim}},\ and\ \bibinfo {author} {\bibfnamefont {Y.~K.}\ \bibnamefont {Semertzidis}},\ }\bibfield  {title} {\bibinfo {title} {{Revisiting the detection rate for axion haloscopes}},\ }\href {https://doi.org/10.1088/1475-7516/2020/03/066} {\bibfield  {journal} {\bibinfo  {journal} {JCAP}\ }\textbf {\bibinfo {volume} {03}},\ \bibinfo {pages} {066}},\ \Eprint {https://arxiv.org/abs/2001.05605} {arXiv:2001.05605 [hep-ex]} \BibitemShut {NoStop}%
\bibitem [{\citenamefont {Bartram}\ \emph {et~al.}(2021)\citenamefont {Bartram}, \citenamefont {Braine}, \citenamefont {Cervantes}, \citenamefont {Crisosto}, \citenamefont {Du}, \citenamefont {Leum}, \citenamefont {Rosenberg}, \citenamefont {Rybka}, \citenamefont {Yang}, \citenamefont {Bowring}, \citenamefont {Chou}, \citenamefont {Khatiwada}, \citenamefont {Sonnenschein}, \citenamefont {Wester}, \citenamefont {Carosi}, \citenamefont {Woollett}, \citenamefont {Duffy}, \citenamefont {Goryachev}, \citenamefont {McAllister}, \citenamefont {Tobar}, \citenamefont {Boutan}, \citenamefont {Jones}, \citenamefont {LaRoque}, \citenamefont {Oblath}, \citenamefont {Taubman}, \citenamefont {Clarke}, \citenamefont {Dove}, \citenamefont {Eddins}, \citenamefont {O'Kelley}, \citenamefont {Nawaz}, \citenamefont {Siddiqi}, \citenamefont {Stevenson}, \citenamefont {Agrawal}, \citenamefont {Dixit}, \citenamefont {Gleason}, \citenamefont {Jois}, \citenamefont {Sikivie}, \citenamefont {Solomon}, \citenamefont {Sullivan},
  \citenamefont {Tanner}, \citenamefont {Lentz}, \citenamefont {Daw}, \citenamefont {Perry}, \citenamefont {Buckley}, \citenamefont {Harrington}, \citenamefont {Henriksen},\ and\ \citenamefont {Murch}}]{PhysRevD.103.032002}%
  \BibitemOpen
  \bibfield  {author} {\bibinfo {author} {\bibfnamefont {C.}~\bibnamefont {Bartram}}, \bibinfo {author} {\bibfnamefont {T.}~\bibnamefont {Braine}}, \bibinfo {author} {\bibfnamefont {R.}~\bibnamefont {Cervantes}}, \bibinfo {author} {\bibfnamefont {N.}~\bibnamefont {Crisosto}}, \bibinfo {author} {\bibfnamefont {N.}~\bibnamefont {Du}}, \bibinfo {author} {\bibfnamefont {G.}~\bibnamefont {Leum}}, \bibinfo {author} {\bibfnamefont {L.~J.}\ \bibnamefont {Rosenberg}}, \bibinfo {author} {\bibfnamefont {G.}~\bibnamefont {Rybka}}, \bibinfo {author} {\bibfnamefont {J.}~\bibnamefont {Yang}}, \bibinfo {author} {\bibfnamefont {D.}~\bibnamefont {Bowring}}, \bibinfo {author} {\bibfnamefont {A.~S.}\ \bibnamefont {Chou}}, \bibinfo {author} {\bibfnamefont {R.}~\bibnamefont {Khatiwada}}, \bibinfo {author} {\bibfnamefont {A.}~\bibnamefont {Sonnenschein}}, \bibinfo {author} {\bibfnamefont {W.}~\bibnamefont {Wester}}, \bibinfo {author} {\bibfnamefont {G.}~\bibnamefont {Carosi}}, \bibinfo {author} {\bibfnamefont {N.}~\bibnamefont
  {Woollett}}, \bibinfo {author} {\bibfnamefont {L.~D.}\ \bibnamefont {Duffy}}, \bibinfo {author} {\bibfnamefont {M.}~\bibnamefont {Goryachev}}, \bibinfo {author} {\bibfnamefont {B.}~\bibnamefont {McAllister}}, \bibinfo {author} {\bibfnamefont {M.~E.}\ \bibnamefont {Tobar}}, \bibinfo {author} {\bibfnamefont {C.}~\bibnamefont {Boutan}}, \bibinfo {author} {\bibfnamefont {M.}~\bibnamefont {Jones}}, \bibinfo {author} {\bibfnamefont {B.~H.}\ \bibnamefont {LaRoque}}, \bibinfo {author} {\bibfnamefont {N.~S.}\ \bibnamefont {Oblath}}, \bibinfo {author} {\bibfnamefont {M.~S.}\ \bibnamefont {Taubman}}, \bibinfo {author} {\bibfnamefont {J.}~\bibnamefont {Clarke}}, \bibinfo {author} {\bibfnamefont {A.}~\bibnamefont {Dove}}, \bibinfo {author} {\bibfnamefont {A.}~\bibnamefont {Eddins}}, \bibinfo {author} {\bibfnamefont {S.~R.}\ \bibnamefont {O'Kelley}}, \bibinfo {author} {\bibfnamefont {S.}~\bibnamefont {Nawaz}}, \bibinfo {author} {\bibfnamefont {I.}~\bibnamefont {Siddiqi}}, \bibinfo {author} {\bibfnamefont
  {N.}~\bibnamefont {Stevenson}}, \bibinfo {author} {\bibfnamefont {A.}~\bibnamefont {Agrawal}}, \bibinfo {author} {\bibfnamefont {A.~V.}\ \bibnamefont {Dixit}}, \bibinfo {author} {\bibfnamefont {J.~R.}\ \bibnamefont {Gleason}}, \bibinfo {author} {\bibfnamefont {S.}~\bibnamefont {Jois}}, \bibinfo {author} {\bibfnamefont {P.}~\bibnamefont {Sikivie}}, \bibinfo {author} {\bibfnamefont {J.~A.}\ \bibnamefont {Solomon}}, \bibinfo {author} {\bibfnamefont {N.~S.}\ \bibnamefont {Sullivan}}, \bibinfo {author} {\bibfnamefont {D.~B.}\ \bibnamefont {Tanner}}, \bibinfo {author} {\bibfnamefont {E.}~\bibnamefont {Lentz}}, \bibinfo {author} {\bibfnamefont {E.~J.}\ \bibnamefont {Daw}}, \bibinfo {author} {\bibfnamefont {M.~G.}\ \bibnamefont {Perry}}, \bibinfo {author} {\bibfnamefont {J.~H.}\ \bibnamefont {Buckley}}, \bibinfo {author} {\bibfnamefont {P.~M.}\ \bibnamefont {Harrington}}, \bibinfo {author} {\bibfnamefont {E.~A.}\ \bibnamefont {Henriksen}},\ and\ \bibinfo {author} {\bibfnamefont {K.~W.}\ \bibnamefont {Murch}}
  (\bibinfo {collaboration} {ADMX Collaboration}),\ }\bibfield  {title} {\bibinfo {title} {Axion dark matter experiment: Run 1b analysis details},\ }\href {https://doi.org/10.1103/PhysRevD.103.032002} {\bibfield  {journal} {\bibinfo  {journal} {Phys. Rev. D}\ }\textbf {\bibinfo {volume} {103}},\ \bibinfo {pages} {032002} (\bibinfo {year} {2021})}\BibitemShut {NoStop}%
\bibitem [{\citenamefont {Yi}\ \emph {et~al.}(2023)\citenamefont {Yi}, \citenamefont {Ahn}, \citenamefont {Kutlu}, \citenamefont {Kim}, \citenamefont {Ko}, \citenamefont {Ivanov}, \citenamefont {Byun}, \citenamefont {van Loo}, \citenamefont {Park}, \citenamefont {Jeong}, \citenamefont {Kwon}, \citenamefont {Nakamura}, \citenamefont {Uchaikin}, \citenamefont {Choi}, \citenamefont {Lee}, \citenamefont {Lee}, \citenamefont {Shin}, \citenamefont {Kim}, \citenamefont {Lee}, \citenamefont {Ahn}, \citenamefont {Bae}, \citenamefont {Lee}, \citenamefont {Kim}, \citenamefont {Gkika}, \citenamefont {Lee}, \citenamefont {Oh}, \citenamefont {Seong}, \citenamefont {Kim}, \citenamefont {Chung}, \citenamefont {Matlashov}, \citenamefont {Youn},\ and\ \citenamefont {Semertzidis}}]{PhysRevLett.130.071002}%
  \BibitemOpen
  \bibfield  {author} {\bibinfo {author} {\bibfnamefont {A.~K.}\ \bibnamefont {Yi}}, \bibinfo {author} {\bibfnamefont {S.}~\bibnamefont {Ahn}}, \bibinfo {author} {\bibfnamefont {i.~m. c. b.~u.}\ \bibnamefont {Kutlu}}, \bibinfo {author} {\bibfnamefont {J.}~\bibnamefont {Kim}}, \bibinfo {author} {\bibfnamefont {B.~R.}\ \bibnamefont {Ko}}, \bibinfo {author} {\bibfnamefont {B.~I.}\ \bibnamefont {Ivanov}}, \bibinfo {author} {\bibfnamefont {H.}~\bibnamefont {Byun}}, \bibinfo {author} {\bibfnamefont {A.~F.}\ \bibnamefont {van Loo}}, \bibinfo {author} {\bibfnamefont {S.}~\bibnamefont {Park}}, \bibinfo {author} {\bibfnamefont {J.}~\bibnamefont {Jeong}}, \bibinfo {author} {\bibfnamefont {O.}~\bibnamefont {Kwon}}, \bibinfo {author} {\bibfnamefont {Y.}~\bibnamefont {Nakamura}}, \bibinfo {author} {\bibfnamefont {S.~V.}\ \bibnamefont {Uchaikin}}, \bibinfo {author} {\bibfnamefont {J.}~\bibnamefont {Choi}}, \bibinfo {author} {\bibfnamefont {S.}~\bibnamefont {Lee}}, \bibinfo {author} {\bibfnamefont {M.}~\bibnamefont {Lee}},
  \bibinfo {author} {\bibfnamefont {Y.~C.}\ \bibnamefont {Shin}}, \bibinfo {author} {\bibfnamefont {J.}~\bibnamefont {Kim}}, \bibinfo {author} {\bibfnamefont {D.}~\bibnamefont {Lee}}, \bibinfo {author} {\bibfnamefont {D.}~\bibnamefont {Ahn}}, \bibinfo {author} {\bibfnamefont {S.}~\bibnamefont {Bae}}, \bibinfo {author} {\bibfnamefont {J.}~\bibnamefont {Lee}}, \bibinfo {author} {\bibfnamefont {Y.}~\bibnamefont {Kim}}, \bibinfo {author} {\bibfnamefont {V.}~\bibnamefont {Gkika}}, \bibinfo {author} {\bibfnamefont {K.~W.}\ \bibnamefont {Lee}}, \bibinfo {author} {\bibfnamefont {S.}~\bibnamefont {Oh}}, \bibinfo {author} {\bibfnamefont {T.}~\bibnamefont {Seong}}, \bibinfo {author} {\bibfnamefont {D.}~\bibnamefont {Kim}}, \bibinfo {author} {\bibfnamefont {W.}~\bibnamefont {Chung}}, \bibinfo {author} {\bibfnamefont {A.}~\bibnamefont {Matlashov}}, \bibinfo {author} {\bibfnamefont {S.}~\bibnamefont {Youn}},\ and\ \bibinfo {author} {\bibfnamefont {Y.~K.}\ \bibnamefont {Semertzidis}},\ }\bibfield  {title} {\bibinfo {title}
  {Axion dark matter search around $4.55\text{ }\text{ }\mathrm{\ensuremath{\mu}}\mathrm{eV}$ with dine-fischler-srednicki-zhitnitskii sensitivity},\ }\href {https://doi.org/10.1103/PhysRevLett.130.071002} {\bibfield  {journal} {\bibinfo  {journal} {Phys. Rev. Lett.}\ }\textbf {\bibinfo {volume} {130}},\ \bibinfo {pages} {071002} (\bibinfo {year} {2023})}\BibitemShut {NoStop}%
\bibitem [{\citenamefont {Stern}\ \emph {et~al.}(2015)\citenamefont {Stern}, \citenamefont {Chisholm}, \citenamefont {Hoskins}, \citenamefont {Sikivie}, \citenamefont {Sullivan}, \citenamefont {Tanner}, \citenamefont {Carosi},\ and\ \citenamefont {van Bibber}}]{Stern:2015kzo}%
  \BibitemOpen
  \bibfield  {author} {\bibinfo {author} {\bibfnamefont {I.}~\bibnamefont {Stern}}, \bibinfo {author} {\bibfnamefont {A.~A.}\ \bibnamefont {Chisholm}}, \bibinfo {author} {\bibfnamefont {J.}~\bibnamefont {Hoskins}}, \bibinfo {author} {\bibfnamefont {P.}~\bibnamefont {Sikivie}}, \bibinfo {author} {\bibfnamefont {N.~S.}\ \bibnamefont {Sullivan}}, \bibinfo {author} {\bibfnamefont {D.~B.}\ \bibnamefont {Tanner}}, \bibinfo {author} {\bibfnamefont {G.}~\bibnamefont {Carosi}},\ and\ \bibinfo {author} {\bibfnamefont {K.}~\bibnamefont {van Bibber}},\ }\bibfield  {title} {\bibinfo {title} {{Cavity design for high-frequency axion dark matter detectors}},\ }\href {https://doi.org/10.1063/1.4938164} {\bibfield  {journal} {\bibinfo  {journal} {Rev. Sci. Instrum.}\ }\textbf {\bibinfo {volume} {86}},\ \bibinfo {pages} {123305} (\bibinfo {year} {2015})},\ \Eprint {https://arxiv.org/abs/1603.06990} {arXiv:1603.06990 [physics.ins-det]} \BibitemShut {NoStop}%
\bibitem [{\citenamefont {Simanovskaia}\ \emph {et~al.}(2021)\citenamefont {Simanovskaia}, \citenamefont {Droster}, \citenamefont {Jackson}, \citenamefont {Urdinaran},\ and\ \citenamefont {van Bibber}}]{10.1063/5.0016125}%
  \BibitemOpen
  \bibfield  {author} {\bibinfo {author} {\bibfnamefont {M.}~\bibnamefont {Simanovskaia}}, \bibinfo {author} {\bibfnamefont {A.}~\bibnamefont {Droster}}, \bibinfo {author} {\bibfnamefont {H.}~\bibnamefont {Jackson}}, \bibinfo {author} {\bibfnamefont {I.}~\bibnamefont {Urdinaran}},\ and\ \bibinfo {author} {\bibfnamefont {K.}~\bibnamefont {van Bibber}},\ }\bibfield  {title} {\bibinfo {title} {{A symmetric multi-rod tunable microwave cavity for a microwave cavity dark matter axion search}},\ }\href {https://doi.org/10.1063/5.0016125} {\bibfield  {journal} {\bibinfo  {journal} {Review of Scientific Instruments}\ }\textbf {\bibinfo {volume} {92}},\ \bibinfo {pages} {033305} (\bibinfo {year} {2021})},\ \Eprint {https://arxiv.org/abs/https://pubs.aip.org/aip/rsi/article-pdf/doi/10.1063/5.0016125/13857698/033305\_1\_online.pdf} {https://pubs.aip.org/aip/rsi/article-pdf/doi/10.1063/5.0016125/13857698/033305\_1\_online.pdf} \BibitemShut {NoStop}%
\bibitem [{\citenamefont {McAllister}\ \emph {et~al.}(2018)\citenamefont {McAllister}, \citenamefont {Flower}, \citenamefont {Tobar},\ and\ \citenamefont {Tobar}}]{PhysRevApplied.9.014028}%
  \BibitemOpen
  \bibfield  {author} {\bibinfo {author} {\bibfnamefont {B.~T.}\ \bibnamefont {McAllister}}, \bibinfo {author} {\bibfnamefont {G.}~\bibnamefont {Flower}}, \bibinfo {author} {\bibfnamefont {L.~E.}\ \bibnamefont {Tobar}},\ and\ \bibinfo {author} {\bibfnamefont {M.~E.}\ \bibnamefont {Tobar}},\ }\bibfield  {title} {\bibinfo {title} {Tunable supermode dielectric resonators for axion dark-matter haloscopes},\ }\href {https://doi.org/10.1103/PhysRevApplied.9.014028} {\bibfield  {journal} {\bibinfo  {journal} {Phys. Rev. Appl.}\ }\textbf {\bibinfo {volume} {9}},\ \bibinfo {pages} {014028} (\bibinfo {year} {2018})}\BibitemShut {NoStop}%
\bibitem [{\citenamefont {Quiskamp}\ \emph {et~al.}(2020)\citenamefont {Quiskamp}, \citenamefont {McAllister}, \citenamefont {Rybka},\ and\ \citenamefont {Tobar}}]{PhysRevApplied.14.044051}%
  \BibitemOpen
  \bibfield  {author} {\bibinfo {author} {\bibfnamefont {A.~P.}\ \bibnamefont {Quiskamp}}, \bibinfo {author} {\bibfnamefont {B.~T.}\ \bibnamefont {McAllister}}, \bibinfo {author} {\bibfnamefont {G.}~\bibnamefont {Rybka}},\ and\ \bibinfo {author} {\bibfnamefont {M.~E.}\ \bibnamefont {Tobar}},\ }\bibfield  {title} {\bibinfo {title} {Dielectric-boosted sensitivity to cylindrical azimuthally varying transverse-magnetic resonant modes in an axion haloscope},\ }\href {https://doi.org/10.1103/PhysRevApplied.14.044051} {\bibfield  {journal} {\bibinfo  {journal} {Phys. Rev. Appl.}\ }\textbf {\bibinfo {volume} {14}},\ \bibinfo {pages} {044051} (\bibinfo {year} {2020})}\BibitemShut {NoStop}%
\bibitem [{\citenamefont {Kim}\ \emph {et~al.}(2020{\natexlab{b}})\citenamefont {Kim}, \citenamefont {Youn}, \citenamefont {Jeong}, \citenamefont {Chung}, \citenamefont {Kwon},\ and\ \citenamefont {Semertzidis}}]{Kim:2019asb}%
  \BibitemOpen
  \bibfield  {author} {\bibinfo {author} {\bibfnamefont {J.}~\bibnamefont {Kim}}, \bibinfo {author} {\bibfnamefont {S.}~\bibnamefont {Youn}}, \bibinfo {author} {\bibfnamefont {J.}~\bibnamefont {Jeong}}, \bibinfo {author} {\bibfnamefont {W.}~\bibnamefont {Chung}}, \bibinfo {author} {\bibfnamefont {O.}~\bibnamefont {Kwon}},\ and\ \bibinfo {author} {\bibfnamefont {Y.~K.}\ \bibnamefont {Semertzidis}},\ }\bibfield  {title} {\bibinfo {title} {{Exploiting higher-order resonant modes for axion haloscopes}},\ }\href {https://doi.org/10.1088/1361-6471/ab5ace} {\bibfield  {journal} {\bibinfo  {journal} {J. Phys. G}\ }\textbf {\bibinfo {volume} {47}},\ \bibinfo {pages} {035203} (\bibinfo {year} {2020}{\natexlab{b}})},\ \Eprint {https://arxiv.org/abs/1910.00793} {arXiv:1910.00793 [physics.ins-det]} \BibitemShut {NoStop}%
\bibitem [{\citenamefont {Alesini}\ \emph {et~al.}(2021)\citenamefont {Alesini} \emph {et~al.}}]{QUAX:2020uxy}%
  \BibitemOpen
  \bibfield  {author} {\bibinfo {author} {\bibfnamefont {D.}~\bibnamefont {Alesini}} \emph {et~al.} (\bibinfo {collaboration} {QUAX}),\ }\bibfield  {title} {\bibinfo {title} {{Realization of a high quality factor resonator with hollow dielectric cylinders for axion searches}},\ }\href {https://doi.org/10.1016/j.nima.2020.164641} {\bibfield  {journal} {\bibinfo  {journal} {Nucl. Instrum. Meth. A}\ }\textbf {\bibinfo {volume} {985}},\ \bibinfo {pages} {164641} (\bibinfo {year} {2021})},\ \Eprint {https://arxiv.org/abs/2004.02754} {arXiv:2004.02754 [physics.ins-det]} \BibitemShut {NoStop}%
\bibitem [{\citenamefont {Di~Vora}\ \emph {et~al.}(2022)\citenamefont {Di~Vora}, \citenamefont {Alesini}, \citenamefont {Braggio}, \citenamefont {Carugno}, \citenamefont {Crescini}, \citenamefont {D'Agostino}, \citenamefont {Di~Gioacchino}, \citenamefont {Falferi}, \citenamefont {Gambardella}, \citenamefont {Gatti}, \citenamefont {Iannone}, \citenamefont {Ligi}, \citenamefont {Lombardi}, \citenamefont {Maccarrone}, \citenamefont {Ortolan}, \citenamefont {Pengo}, \citenamefont {Rettaroli}, \citenamefont {Ruoso}, \citenamefont {Taffarello},\ and\ \citenamefont {Tocci}}]{PhysRevApplied.17.054013}%
  \BibitemOpen
  \bibfield  {author} {\bibinfo {author} {\bibfnamefont {R.}~\bibnamefont {Di~Vora}}, \bibinfo {author} {\bibfnamefont {D.}~\bibnamefont {Alesini}}, \bibinfo {author} {\bibfnamefont {C.}~\bibnamefont {Braggio}}, \bibinfo {author} {\bibfnamefont {G.}~\bibnamefont {Carugno}}, \bibinfo {author} {\bibfnamefont {N.}~\bibnamefont {Crescini}}, \bibinfo {author} {\bibfnamefont {D.}~\bibnamefont {D'Agostino}}, \bibinfo {author} {\bibfnamefont {D.}~\bibnamefont {Di~Gioacchino}}, \bibinfo {author} {\bibfnamefont {P.}~\bibnamefont {Falferi}}, \bibinfo {author} {\bibfnamefont {U.}~\bibnamefont {Gambardella}}, \bibinfo {author} {\bibfnamefont {C.}~\bibnamefont {Gatti}}, \bibinfo {author} {\bibfnamefont {G.}~\bibnamefont {Iannone}}, \bibinfo {author} {\bibfnamefont {C.}~\bibnamefont {Ligi}}, \bibinfo {author} {\bibfnamefont {A.}~\bibnamefont {Lombardi}}, \bibinfo {author} {\bibfnamefont {G.}~\bibnamefont {Maccarrone}}, \bibinfo {author} {\bibfnamefont {A.}~\bibnamefont {Ortolan}}, \bibinfo {author} {\bibfnamefont
  {R.}~\bibnamefont {Pengo}}, \bibinfo {author} {\bibfnamefont {A.}~\bibnamefont {Rettaroli}}, \bibinfo {author} {\bibfnamefont {G.}~\bibnamefont {Ruoso}}, \bibinfo {author} {\bibfnamefont {L.}~\bibnamefont {Taffarello}},\ and\ \bibinfo {author} {\bibfnamefont {S.}~\bibnamefont {Tocci}},\ }\bibfield  {title} {\bibinfo {title} {High-$q$ microwave dielectric resonator for axion dark-matter haloscopes},\ }\href {https://doi.org/10.1103/PhysRevApplied.17.054013} {\bibfield  {journal} {\bibinfo  {journal} {Phys. Rev. Appl.}\ }\textbf {\bibinfo {volume} {17}},\ \bibinfo {pages} {054013} (\bibinfo {year} {2022})}\BibitemShut {NoStop}%
\bibitem [{\citenamefont {Alesini}\ \emph {et~al.}(2020)\citenamefont {Alesini} \emph {et~al.}}]{QUAX:2020wfd}%
  \BibitemOpen
  \bibfield  {author} {\bibinfo {author} {\bibfnamefont {D.}~\bibnamefont {Alesini}} \emph {et~al.} (\bibinfo {collaboration} {QUAX}),\ }\bibfield  {title} {\bibinfo {title} {{High quality factor photonic cavity for dark matter axion searches}},\ }\href {https://doi.org/10.1063/5.0003878} {\bibfield  {journal} {\bibinfo  {journal} {Rev. Sci. Instrum.}\ }\textbf {\bibinfo {volume} {91}},\ \bibinfo {pages} {094701} (\bibinfo {year} {2020})},\ \Eprint {https://arxiv.org/abs/2002.01816} {arXiv:2002.01816 [physics.ins-det]} \BibitemShut {NoStop}%
\bibitem [{\citenamefont {Bae}\ \emph {et~al.}(2023)\citenamefont {Bae}, \citenamefont {Youn},\ and\ \citenamefont {Jeong}}]{PhysRevD.107.015012}%
  \BibitemOpen
  \bibfield  {author} {\bibinfo {author} {\bibfnamefont {S.}~\bibnamefont {Bae}}, \bibinfo {author} {\bibfnamefont {S.}~\bibnamefont {Youn}},\ and\ \bibinfo {author} {\bibfnamefont {J.}~\bibnamefont {Jeong}},\ }\bibfield  {title} {\bibinfo {title} {Tunable photonic crystal haloscope for high-mass axion searches},\ }\href {https://doi.org/10.1103/PhysRevD.107.015012} {\bibfield  {journal} {\bibinfo  {journal} {Phys. Rev. D}\ }\textbf {\bibinfo {volume} {107}},\ \bibinfo {pages} {015012} (\bibinfo {year} {2023})}\BibitemShut {NoStop}%
\bibitem [{\citenamefont {Kishimoto}\ \emph {et~al.}(2021)\citenamefont {Kishimoto}, \citenamefont {Suzuki}, \citenamefont {Ogawa}, \citenamefont {Mori},\ and\ \citenamefont {Yamashita}}]{Kishimoto:2021ral}%
  \BibitemOpen
  \bibfield  {author} {\bibinfo {author} {\bibfnamefont {Y.}~\bibnamefont {Kishimoto}}, \bibinfo {author} {\bibfnamefont {Y.}~\bibnamefont {Suzuki}}, \bibinfo {author} {\bibfnamefont {I.}~\bibnamefont {Ogawa}}, \bibinfo {author} {\bibfnamefont {Y.}~\bibnamefont {Mori}},\ and\ \bibinfo {author} {\bibfnamefont {M.}~\bibnamefont {Yamashita}},\ }\bibfield  {title} {\bibinfo {title} {{Development of a cavity with photonic crystal structure for axion searches}},\ }\href {https://doi.org/10.1093/ptep/ptab051} {\bibfield  {journal} {\bibinfo  {journal} {PTEP}\ }\textbf {\bibinfo {volume} {2021}},\ \bibinfo {pages} {063H01} (\bibinfo {year} {2021})},\ \Eprint {https://arxiv.org/abs/2103.00101} {arXiv:2103.00101 [hep-ex]} \BibitemShut {NoStop}%
\bibitem [{\citenamefont {Lawson}\ \emph {et~al.}(2019)\citenamefont {Lawson}, \citenamefont {Millar}, \citenamefont {Pancaldi}, \citenamefont {Vitagliano},\ and\ \citenamefont {Wilczek}}]{PhysRevLett.123.141802}%
  \BibitemOpen
  \bibfield  {author} {\bibinfo {author} {\bibfnamefont {M.}~\bibnamefont {Lawson}}, \bibinfo {author} {\bibfnamefont {A.~J.}\ \bibnamefont {Millar}}, \bibinfo {author} {\bibfnamefont {M.}~\bibnamefont {Pancaldi}}, \bibinfo {author} {\bibfnamefont {E.}~\bibnamefont {Vitagliano}},\ and\ \bibinfo {author} {\bibfnamefont {F.}~\bibnamefont {Wilczek}},\ }\bibfield  {title} {\bibinfo {title} {Tunable axion plasma haloscopes},\ }\href {https://doi.org/10.1103/PhysRevLett.123.141802} {\bibfield  {journal} {\bibinfo  {journal} {Phys. Rev. Lett.}\ }\textbf {\bibinfo {volume} {123}},\ \bibinfo {pages} {141802} (\bibinfo {year} {2019})}\BibitemShut {NoStop}%
\bibitem [{\citenamefont {Millar}\ \emph {et~al.}(2023)\citenamefont {Millar} \emph {et~al.}}]{ALPHA:2022rxj}%
  \BibitemOpen
  \bibfield  {author} {\bibinfo {author} {\bibfnamefont {A.~J.}\ \bibnamefont {Millar}} \emph {et~al.} (\bibinfo {collaboration} {ALPHA}),\ }\bibfield  {title} {\bibinfo {title} {{Searching for dark matter with plasma haloscopes}},\ }\href {https://doi.org/10.1103/PhysRevD.107.055013} {\bibfield  {journal} {\bibinfo  {journal} {Phys. Rev. D}\ }\textbf {\bibinfo {volume} {107}},\ \bibinfo {pages} {055013} (\bibinfo {year} {2023})},\ \Eprint {https://arxiv.org/abs/2210.00017} {arXiv:2210.00017 [hep-ph]} \BibitemShut {NoStop}%
\bibitem [{\citenamefont {Jeong}\ \emph {et~al.}(2018{\natexlab{a}})\citenamefont {Jeong}, \citenamefont {Youn}, \citenamefont {Ahn}, \citenamefont {Kim},\ and\ \citenamefont {Semertzidis}}]{Jeong:2017hqs}%
  \BibitemOpen
  \bibfield  {author} {\bibinfo {author} {\bibfnamefont {J.}~\bibnamefont {Jeong}}, \bibinfo {author} {\bibfnamefont {S.}~\bibnamefont {Youn}}, \bibinfo {author} {\bibfnamefont {S.}~\bibnamefont {Ahn}}, \bibinfo {author} {\bibfnamefont {J.~E.}\ \bibnamefont {Kim}},\ and\ \bibinfo {author} {\bibfnamefont {Y.~K.}\ \bibnamefont {Semertzidis}},\ }\bibfield  {title} {\bibinfo {title} {{Concept of multiple-cell cavity for axion dark matter search}},\ }\href {https://doi.org/10.1016/j.physletb.2017.12.066} {\bibfield  {journal} {\bibinfo  {journal} {Phys. Lett. B}\ }\textbf {\bibinfo {volume} {777}},\ \bibinfo {pages} {412} (\bibinfo {year} {2018}{\natexlab{a}})},\ \Eprint {https://arxiv.org/abs/1710.06969} {arXiv:1710.06969 [astro-ph.IM]} \BibitemShut {NoStop}%
\bibitem [{\citenamefont {Jeong}\ \emph {et~al.}(2023)\citenamefont {Jeong}, \citenamefont {Youn},\ and\ \citenamefont {Kim}}]{Jeong:2022akg}%
  \BibitemOpen
  \bibfield  {author} {\bibinfo {author} {\bibfnamefont {J.}~\bibnamefont {Jeong}}, \bibinfo {author} {\bibfnamefont {S.}~\bibnamefont {Youn}},\ and\ \bibinfo {author} {\bibfnamefont {J.~E.}\ \bibnamefont {Kim}},\ }\bibfield  {title} {\bibinfo {title} {{Multiple-cell cavity design for high mass axion searches: An extended study}},\ }\href {https://doi.org/10.1016/j.nima.2023.168327} {\bibfield  {journal} {\bibinfo  {journal} {Nucl. Instrum. Meth. A}\ }\textbf {\bibinfo {volume} {1053}},\ \bibinfo {pages} {168327} (\bibinfo {year} {2023})},\ \Eprint {https://arxiv.org/abs/2205.01319} {arXiv:2205.01319 [hep-ex]} \BibitemShut {NoStop}%
\bibitem [{\citenamefont {\'Alvarez~Melc\'on}\ \emph {et~al.}(2020)\citenamefont {\'Alvarez~Melc\'on} \emph {et~al.}}]{AlvarezMelcon:2020vee}%
  \BibitemOpen
  \bibfield  {author} {\bibinfo {author} {\bibfnamefont {A.}~\bibnamefont {\'Alvarez~Melc\'on}} \emph {et~al.},\ }\bibfield  {title} {\bibinfo {title} {{Scalable haloscopes for axion dark matter detection in the 30$\mu$eV range with RADES}},\ }\href {https://doi.org/10.1007/JHEP07(2020)084} {\bibfield  {journal} {\bibinfo  {journal} {JHEP}\ }\textbf {\bibinfo {volume} {07}},\ \bibinfo {pages} {084}},\ \Eprint {https://arxiv.org/abs/2002.07639} {arXiv:2002.07639 [hep-ex]} \BibitemShut {NoStop}%
\bibitem [{\citenamefont {Hagmann}\ \emph {et~al.}(1990)\citenamefont {Hagmann}, \citenamefont {Sikivie}, \citenamefont {Sullivan}, \citenamefont {Tanner},\ and\ \citenamefont {Cho}}]{10.1063/1.1141427}%
  \BibitemOpen
  \bibfield  {author} {\bibinfo {author} {\bibfnamefont {C.}~\bibnamefont {Hagmann}}, \bibinfo {author} {\bibfnamefont {P.}~\bibnamefont {Sikivie}}, \bibinfo {author} {\bibfnamefont {N.}~\bibnamefont {Sullivan}}, \bibinfo {author} {\bibfnamefont {D.~B.}\ \bibnamefont {Tanner}},\ and\ \bibinfo {author} {\bibfnamefont {S.}~\bibnamefont {Cho}},\ }\bibfield  {title} {\bibinfo {title} {{Cavity design for a cosmic axion detector}},\ }\href {https://doi.org/10.1063/1.1141427} {\bibfield  {journal} {\bibinfo  {journal} {Review of Scientific Instruments}\ }\textbf {\bibinfo {volume} {61}},\ \bibinfo {pages} {1076} (\bibinfo {year} {1990})},\ \Eprint {https://arxiv.org/abs/https://pubs.aip.org/aip/rsi/article-pdf/61/3/1076/19023464/1076\_1\_online.pdf} {https://pubs.aip.org/aip/rsi/article-pdf/61/3/1076/19023464/1076\_1\_online.pdf} \BibitemShut {NoStop}%
\bibitem [{\citenamefont {Kinion}(2001{\natexlab{a}})}]{Kinion:2001fp}%
  \BibitemOpen
  \bibfield  {author} {\bibinfo {author} {\bibfnamefont {D.~S.}\ \bibnamefont {Kinion}},\ }\emph {\bibinfo {title} {{First results from a multiple microwave cavity search for dark matter axions}}},\ \href@noop {} {\bibinfo {type} {Other thesis}} (\bibinfo {year} {2001}{\natexlab{a}})\BibitemShut {NoStop}%
\bibitem [{\citenamefont {Yang}\ \emph {et~al.}(2020)\citenamefont {Yang}, \citenamefont {Gleason}, \citenamefont {Jois}, \citenamefont {Stern}, \citenamefont {Sikivie}, \citenamefont {Sullivan},\ and\ \citenamefont {Tanner}}]{Yang:2020xsc}%
  \BibitemOpen
  \bibfield  {author} {\bibinfo {author} {\bibfnamefont {J.}~\bibnamefont {Yang}}, \bibinfo {author} {\bibfnamefont {J.~R.}\ \bibnamefont {Gleason}}, \bibinfo {author} {\bibfnamefont {S.}~\bibnamefont {Jois}}, \bibinfo {author} {\bibfnamefont {I.}~\bibnamefont {Stern}}, \bibinfo {author} {\bibfnamefont {P.}~\bibnamefont {Sikivie}}, \bibinfo {author} {\bibfnamefont {N.~S.}\ \bibnamefont {Sullivan}},\ and\ \bibinfo {author} {\bibfnamefont {D.~B.}\ \bibnamefont {Tanner}},\ }\bibfield  {title} {\bibinfo {title} {{Search for 5\textendash{}9 $\mu$eV Axions with ADMX Four-Cavity Array}},\ }\href {https://doi.org/10.1007/978-3-030-43761-9_7} {\bibfield  {journal} {\bibinfo  {journal} {Springer Proc. Phys.}\ }\textbf {\bibinfo {volume} {245}},\ \bibinfo {pages} {53} (\bibinfo {year} {2020})}\BibitemShut {NoStop}%
\bibitem [{\citenamefont {Adair}\ \emph {et~al.}(2022)\citenamefont {Adair} \emph {et~al.}}]{Adair:2022rtw}%
  \BibitemOpen
  \bibfield  {author} {\bibinfo {author} {\bibfnamefont {C.~M.}\ \bibnamefont {Adair}} \emph {et~al.},\ }\bibfield  {title} {\bibinfo {title} {{Search for Dark Matter Axions with CAST-CAPP}},\ }\href {https://doi.org/10.1038/s41467-022-33913-6} {\bibfield  {journal} {\bibinfo  {journal} {Nature Commun.}\ }\textbf {\bibinfo {volume} {13}},\ \bibinfo {pages} {6180} (\bibinfo {year} {2022})},\ \Eprint {https://arxiv.org/abs/2211.02902} {arXiv:2211.02902 [hep-ex]} \BibitemShut {NoStop}%
\bibitem [{\citenamefont {Jeong}\ \emph {et~al.}(2018{\natexlab{b}})\citenamefont {Jeong}, \citenamefont {Youn}, \citenamefont {Ahn}, \citenamefont {Kang},\ and\ \citenamefont {Semertzidis}}]{JEONG201833}%
  \BibitemOpen
  \bibfield  {author} {\bibinfo {author} {\bibfnamefont {J.}~\bibnamefont {Jeong}}, \bibinfo {author} {\bibfnamefont {S.}~\bibnamefont {Youn}}, \bibinfo {author} {\bibfnamefont {S.}~\bibnamefont {Ahn}}, \bibinfo {author} {\bibfnamefont {C.}~\bibnamefont {Kang}},\ and\ \bibinfo {author} {\bibfnamefont {Y.~K.}\ \bibnamefont {Semertzidis}},\ }\bibfield  {title} {\bibinfo {title} {Phase-matching of multiple-cavity detectors for dark matter axion search},\ }\href {https://doi.org/https://doi.org/10.1016/j.astropartphys.2017.10.012} {\bibfield  {journal} {\bibinfo  {journal} {Astroparticle Physics}\ }\textbf {\bibinfo {volume} {97}},\ \bibinfo {pages} {33} (\bibinfo {year} {2018}{\natexlab{b}})}\BibitemShut {NoStop}%
\bibitem [{\citenamefont {Kinion}(2001{\natexlab{b}})}]{kinion2001first}%
  \BibitemOpen
  \bibfield  {author} {\bibinfo {author} {\bibfnamefont {D.~S.}\ \bibnamefont {Kinion}},\ }\href@noop {} {\emph {\bibinfo {title} {First results from a multiple-microwave-cavity search for dark-matter axions}}}\ (\bibinfo  {publisher} {University of California, Davis},\ \bibinfo {year} {2001})\BibitemShut {NoStop}%
\bibitem [{\citenamefont {Kuo}(2020)}]{Kuo:2019cps}%
  \BibitemOpen
  \bibfield  {author} {\bibinfo {author} {\bibfnamefont {C.-L.}\ \bibnamefont {Kuo}},\ }\bibfield  {title} {\bibinfo {title} {{Large-Volume Centimeter-Wave Cavities for Axion Searches}},\ }\href {https://doi.org/10.1088/1475-7516/2020/06/010} {\bibfield  {journal} {\bibinfo  {journal} {JCAP}\ }\textbf {\bibinfo {volume} {06}},\ \bibinfo {pages} {010}},\ \Eprint {https://arxiv.org/abs/1910.04156} {arXiv:1910.04156 [physics.ins-det]} \BibitemShut {NoStop}%
\bibitem [{\citenamefont {Kuo}(2021)}]{Kuo:2020llc}%
  \BibitemOpen
  \bibfield  {author} {\bibinfo {author} {\bibfnamefont {C.-L.}\ \bibnamefont {Kuo}},\ }\bibfield  {title} {\bibinfo {title} {{Symmetrically Tuned Large-Volume Conic Shell-Cavities for Axion Searches}},\ }\href {https://doi.org/10.1088/1475-7516/2021/02/018} {\bibfield  {journal} {\bibinfo  {journal} {JCAP}\ }\textbf {\bibinfo {volume} {02}},\ \bibinfo {pages} {018}},\ \Eprint {https://arxiv.org/abs/2010.04337} {arXiv:2010.04337 [physics.ins-det]} \BibitemShut {NoStop}%
\bibitem [{\citenamefont {Dyson}\ \emph {et~al.}(2024)\citenamefont {Dyson}, \citenamefont {Bartram}, \citenamefont {Davidson}, \citenamefont {Ezekiel}, \citenamefont {Futamura}, \citenamefont {Liu},\ and\ \citenamefont {Kuo}}]{Dyson:2024elo}%
  \BibitemOpen
  \bibfield  {author} {\bibinfo {author} {\bibfnamefont {T.~A.}\ \bibnamefont {Dyson}}, \bibinfo {author} {\bibfnamefont {C.~L.}\ \bibnamefont {Bartram}}, \bibinfo {author} {\bibfnamefont {A.}~\bibnamefont {Davidson}}, \bibinfo {author} {\bibfnamefont {J.~B.}\ \bibnamefont {Ezekiel}}, \bibinfo {author} {\bibfnamefont {L.~M.}\ \bibnamefont {Futamura}}, \bibinfo {author} {\bibfnamefont {T.}~\bibnamefont {Liu}},\ and\ \bibinfo {author} {\bibfnamefont {C.-L.}\ \bibnamefont {Kuo}},\ }\bibfield  {title} {\bibinfo {title} {{Demonstration of a high-volume tunable axion haloscope above 7 GHz}},\ }\href@noop {} {\  (\bibinfo {year} {2024})},\ \Eprint {https://arxiv.org/abs/2402.01060} {arXiv:2402.01060 [physics.ins-det]} \BibitemShut {NoStop}%
\bibitem [{\citenamefont {Bradley}\ \emph {et~al.}(2003)\citenamefont {Bradley}, \citenamefont {Clarke}, \citenamefont {Kinion}, \citenamefont {Rosenberg}, \citenamefont {van Bibber}, \citenamefont {Matsuki}, \citenamefont {M\"uck},\ and\ \citenamefont {Sikivie}}]{RevModPhys.75.777}%
  \BibitemOpen
  \bibfield  {author} {\bibinfo {author} {\bibfnamefont {R.}~\bibnamefont {Bradley}}, \bibinfo {author} {\bibfnamefont {J.}~\bibnamefont {Clarke}}, \bibinfo {author} {\bibfnamefont {D.}~\bibnamefont {Kinion}}, \bibinfo {author} {\bibfnamefont {L.~J.}\ \bibnamefont {Rosenberg}}, \bibinfo {author} {\bibfnamefont {K.}~\bibnamefont {van Bibber}}, \bibinfo {author} {\bibfnamefont {S.}~\bibnamefont {Matsuki}}, \bibinfo {author} {\bibfnamefont {M.}~\bibnamefont {M\"uck}},\ and\ \bibinfo {author} {\bibfnamefont {P.}~\bibnamefont {Sikivie}},\ }\bibfield  {title} {\bibinfo {title} {Microwave cavity searches for dark-matter axions},\ }\href {https://doi.org/10.1103/RevModPhys.75.777} {\bibfield  {journal} {\bibinfo  {journal} {Rev. Mod. Phys.}\ }\textbf {\bibinfo {volume} {75}},\ \bibinfo {pages} {777} (\bibinfo {year} {2003})}\BibitemShut {NoStop}%
\bibitem [{Note1()}]{Note1}%
  \BibitemOpen
  \bibinfo {note} {Nonetheless, provided each cell has its own tunable antenna, the present polygonal resonator becomes a cavity array in which the multiple cavities have $C\sim 0.6$, $Q\sim 60000$ and are all tuned together over a large range. In addition, for phase-matched readout, the cavity cells would only need an additional fine-tuning mechanism to compensate for small frequency differences between cells.}\BibitemShut {Stop}%
\bibitem [{\citenamefont {Som}\ \emph {et~al.}(2011)\citenamefont {Som}, \citenamefont {Seth}, \citenamefont {Mandal},\ and\ \citenamefont {Ghosh}}]{som2011bead}%
  \BibitemOpen
  \bibfield  {author} {\bibinfo {author} {\bibfnamefont {S.}~\bibnamefont {Som}}, \bibinfo {author} {\bibfnamefont {S.}~\bibnamefont {Seth}}, \bibinfo {author} {\bibfnamefont {A.}~\bibnamefont {Mandal}},\ and\ \bibinfo {author} {\bibfnamefont {S.}~\bibnamefont {Ghosh}},\ }\bibfield  {title} {\bibinfo {title} {Bead-pull measurement using phase-shift technique in multi-cell elliptical cavity},\ }\href@noop {} {\bibfield  {journal} {\bibinfo  {journal} {Proceedings of IPAC}\ }\textbf {\bibinfo {volume} {280}},\ \bibinfo {pages} {2011} (\bibinfo {year} {2011})}\BibitemShut {NoStop}%
\bibitem [{\citenamefont {Ahn}\ \emph {et~al.}(2022)\citenamefont {Ahn}, \citenamefont {Kwon}, \citenamefont {Chung}, \citenamefont {Jang}, \citenamefont {Lee}, \citenamefont {Lee}, \citenamefont {Youn}, \citenamefont {Byun}, \citenamefont {Youm},\ and\ \citenamefont {Semertzidis}}]{PhysRevApplied.17.L061005}%
  \BibitemOpen
  \bibfield  {author} {\bibinfo {author} {\bibfnamefont {D.}~\bibnamefont {Ahn}}, \bibinfo {author} {\bibfnamefont {O.}~\bibnamefont {Kwon}}, \bibinfo {author} {\bibfnamefont {W.}~\bibnamefont {Chung}}, \bibinfo {author} {\bibfnamefont {W.}~\bibnamefont {Jang}}, \bibinfo {author} {\bibfnamefont {D.}~\bibnamefont {Lee}}, \bibinfo {author} {\bibfnamefont {J.}~\bibnamefont {Lee}}, \bibinfo {author} {\bibfnamefont {S.~W.}\ \bibnamefont {Youn}}, \bibinfo {author} {\bibfnamefont {H.}~\bibnamefont {Byun}}, \bibinfo {author} {\bibfnamefont {D.}~\bibnamefont {Youm}},\ and\ \bibinfo {author} {\bibfnamefont {Y.~K.}\ \bibnamefont {Semertzidis}},\ }\bibfield  {title} {\bibinfo {title} {Biaxially textured ${\mathrm{yba}}_{2}{\mathrm{cu}}_{3}{\mathrm{o}}_{7\ensuremath{-}x}$ microwave cavity in a high magnetic field for a dark-matter axion search},\ }\href {https://doi.org/10.1103/PhysRevApplied.17.L061005} {\bibfield  {journal} {\bibinfo  {journal} {Phys. Rev. Appl.}\ }\textbf {\bibinfo {volume} {17}},\ \bibinfo {pages}
  {L061005} (\bibinfo {year} {2022})}\BibitemShut {NoStop}%
\end{thebibliography}%

\end{document}